\documentclass[a4paper,fleqn,usenatbib]{mnras}

\usepackage{newtxtext,newtxmath}

\usepackage[T1]{fontenc}
\usepackage{ae,aecompl}
\usepackage{relsize} 
\usepackage{color,soul}
\usepackage{subfig}

\usepackage{graphicx}
\usepackage{amsmath}
\usepackage{amssymb}
\usepackage{bm}
\usepackage{footnote}
\usepackage{bigints}

\newcommand{\msun}{\mbox{M$_\odot$}}

\newcommand{\pc}{\mbox{${\rm pc}$}}

\title[Stellar birth environment \& PPD dispersal]{\vspace{-2mm}Prevalent externally-driven protoplanetary disc dispersal as a function of the galactic environment\vspace{-3mm}}

\author[A.~J.~Winter et al.]{Andrew~J.~Winter,$^{1,2, 3}$\thanks{ajwinter@ast.cam.ac.uk}	
 J.~M.~Diederik~Kruijssen,$^1$ M\'{e}lanie Chevance,$^1$ \newauthor Benjamin W. Keller,$^{1}$ Steven N. Longmore$^{4}$\\
$^{1}$Astronomisches Rechen-Institut, Zentrum f\"{u}r Astronomie der Universit\"{a}t Heidelberg, M\"{o}nchhofstra\ss e 12-14, 69120 Heidelberg, Germany \\
$^{2}$Institute of Astronomy, University of Cambridge, Madingley Road, Cambridge CB3 0HA, UK \\
$^{3}$Department of Physics and Astronomy, University of Leicester, Leicester, LE1 7RH, UK\\
$^{4}$Astrophysics Research Institute, Liverpool John Moores University, IC2, Liverpool Science Park, 146 Brownlow Hill, Liverpool L3 5RF, UK\vspace{-3mm}
}

\date{Accepted 2019 September 27; Received 2019 September 23; in original form 2019 July 4}\vspace{-2mm}

\pubyear{2019}

\begin{document}
\label{firstpage}
\pagerange{\pageref{firstpage}--\pageref{lastpage}}
\maketitle

\begin{abstract}
The stellar birth environment can significantly shorten protoplanetary disc (PPD) lifetimes due to the influence of stellar feedback mechanisms. The degree to which these mechanisms suppress the time and mass available for planet formation is dependent on the local far-ultraviolet (FUV) field strength, stellar density, and ISM properties. In this work, we present the first theoretical framework quantifying the distribution of PPD dispersal time-scales as a function of parameters that describe the galactic environment. We calculate the probability density function for FUV flux and stellar density in the solar neighbourhood. In agreement with previous studies, we find that external photoevaporation is the dominant environment-related factor influencing local stellar populations after the embedded phase. Applying our general prescription to the Central Molecular Zone of the Milky Way (i.e.\ the central $\sim250~\pc$), we predict that $90\%$ of PPDs in the region are destroyed within $1$~Myr of the dispersal of the parent molecular cloud. Even in such dense environments, we find that external photoevaporation is the dominant disc depletion mechanism over dynamical encounters between stars. PPDs around low-mass stars are particularly sensitive to FUV-induced mass loss, due to a shallower gravitational potential. For stars of mass $\sim 1~\msun$, the solar neighbourhood lies at approximately the highest gas surface density for which PPD dispersal is still relatively unaffected by external FUV photons, with a median PPD dispersal timescale of $\sim 4$~Myr. We highlight the key questions to be addressed to further contextualise the significance of the local galactic environment for planet formation.
\end{abstract}

\begin{keywords} 
planets and satellites: formation --- protoplanetary discs --- stars: formation --- galaxies: ISM --- galaxies: star clusters: general --- galaxies: star formation\vspace{-2mm}
\end{keywords}


\section{Introduction}

The process of planet formation is strongly dependent on the stellar birth environment. The majority of stars exist in clusters or associations within their first few Myr of evolution \citep{Lad03, Lon14, Krum19}, during which time they also host protoplanetary discs \citep[PPDs - e.g.][]{Hai01b, Rib14}. Multiple feedback mechanisms influence disc evolution. In sufficiently dense environments, star-disc encounters can truncate the disc and induce increased accretion rates \citep{Cla93, Ost94, Hal96, Pfa05, Olc06, Pfa06, dJO12, Bre14, Ros14, Win18}. Recent studies indicate that in the solar neighbourhood such interactions only have a significant effect in the early stages of cluster evolution due to enhanced stellar multiplicity and substructure, and therefore set initial conditions rather than destruction time-scales \citep{Win18b, Win18c, Bat18}. However, in regions with massive stars, external photoevaporation by far-ultraviolet (FUV) and extreme-ultraviolet (EUV) photons can rapidly disperse PPDs \citep{Joh98, Sto99,Arm00,Cla07, Fat08,Ada10,Fac16,Ans17,Haw18b, Win18b}. Additionally, before the dispersal of the parent giant molecular cloud (GMC), ram pressure stripping can truncate PPDs \citep{Wij17} or additional material can be accreted \citep{Moe09, Sci14}, leading to the destruction and reforming of discs during the embedded phase \citep{Bat18}. If a PPD is destroyed quickly by feedback in dense stellar environments, planets may be unable to form, depending on the efficiency of the formation mechanisms \citep{You05, Joh17, Orm17, Haw18}. Given the apparent ubiquity of grouped star formation, quantifying the destruction time-scales for PPDs due to neighbour feedback is of great relevance for understanding the demographics of PPDs and exoplanetary systems. 

Although stars are understood to form primarily in groups, the nature of those groups is diverse and remains the topic of debate; however, it is evident that a density continuum well describes the distribution of the interstellar medium (\citealt{Vaz94}; \citealt{Pad02}; \mbox{\citealt{Hil12}}) and stars \citep{Bre10, Kru12}. Previous statistical investigations into PPD destruction by stellar feedback have been focused on young star-forming environments in the solar neighbourhood ($\lesssim 1$--$2$~kpc from the Sun, e.g.\ \citealt{Fat08}). However, this approach may not yield a representative picture, since star and planet formation in the Milky Way historically proceeded at much greater gas densities, similar to those seen near the galactic centre \citep{Kru13}. 

Recent work shows that the properties of GMCs and young stellar clusters depend on the galactic-scale interstellar medium (ISM) properties \citep[e.g.][]{Bol08, Hey09, Lon14, Ada15, Fre17, Rei17, Sun18}. For example, the density threshold required for star formation to proceed is at least an order of magnitude higher in the central $\sim250~\pc$ of the Milky Way (the Central Molecular Zone; CMZ) than in the solar neighbourhood \citep{Lon13, Kru14, Rat14, Gin18}. Given that at higher densities, star-disc encounters are more frequent and FUV fields are stronger \citep{Win18b}, we would expect PPD lifetimes to be reduced in the CMZ with respect to the galactic disc. Indeed, preliminary studies into the PPD population towards the CMZ indicate low disc survival fractions in young stellar populations \citep{Sto10, Sto15}.

Due to the above considerations, this work is aimed at linking PPD lifetimes to the distribution of molecular gas from which the stellar populations form, thereby establishing time-scales available for planet formation as a function of quantities describing the local galactic environment. We will primarily consider the influence of FUV-induced mass loss, which \citet{Win18b} demonstrates to dominate over dynamical encounters in observed environments.\footnote{As discussed previously, ram pressure stripping and dynamical encounters can also alter disc evolution, and they are further discussed in Section~\ref{sec:ppd_dest}.} This represents a generalisation of the study of \citet{Fat08}, where only local star-forming environments were considered. The present study also incorporates other recent advances in our understanding of GMC properties and clustered star formation. Due to recent developments in the theory of FUV-induced PPD mass loss rates \citep{Fac16,Haw18b}, we are additionally able to estimate the time-scales for PPD destruction based on the properties of the disc and the mass of its host star. 

In this paper, we first review the time-scales for PPD dispersal as a result of environmental influences in Section~\ref{sec:ppd_dest}. We then characterise the stellar birth environment by establishing probability density functions (PDFs) in stellar density--FUV flux space as a function of the average gas properties in the solar neighbourhood and the CMZ (Section~\ref{sec:stellar_birth}). For the reader interested only in our main results, Section~\ref{sec:discuss} discusses PPD lifetimes as a function of large scale ISM properties. We present our concluding remarks in Section~\ref{sec:concs}.

\section{PPD destruction time-scales}
\label{sec:ppd_dest}

\subsection{External photoevaporation}

In this section, we will establish characteristic time-scales for PPD destruction due to FUV irradiation. We will henceforth disregard the mass loss of PPDs due to EUV flux  because photons in that energy range only dominate mass loss at extremely small ($\ll 0.1$~pc) and large ($\gg 10$~pc) spatial separations from massive stars \citep[although these numbers depend on disc properties and the irradiating source --][]{Sto99, Win18b}. In this case, time-scales for disc dispersal by external photoevaporation are either rapid ($\ll 1$~Myr) or slow ($\gg 3$~Myr) respectively. For rapid dispersal we are less interested in establishing the exact time at which PPDs are destroyed, and more in the prediction that they are sufficiently short-lived such that planet formation is likely suppressed or significantly influenced. For discs where the FUV-induced mass loss is small, PPD depletion is dictated by internal processes such as internal photoevaporation and accretion. We therefore limit our attention to the influence of FUV photons in the following discussion. Henceforth we will refer to the flux $F$ and luminosity $L$ without subscripts for simplicity; it is to be understood that we refer only to the contribution of the FUV photons.

We will now calculate the FUV-induced PPD destruction time-scale $\tau_\mathrm{FUV}$ as a function of FUV flux $F$, host mass $m_*$ and viscous time-scale $\tau_\mathrm{visc}$. Each of these parameters has a significant impact on survival time-scales. FUV mass loss rates increase with $F$ for $F\lesssim 10^4 \, G_0$; above this threshold the temperature of the photodissociation region is a weak function of $F$, and therefore so too is the mass flux in the thermal wind \citep[e.g.][]{Tie85, Hol97, Joh98}. The efficiency of FUV-induced mass loss increases with decreasing stellar host mass mass simply due to a shallower gravitational potential, and therefore lower escape velocity. Finally, the time-scale for viscous spreading is important since this dictates the rate at which material is accreted onto the central star, and the dispersal time-scale once the reservoir of material in outer disc has been depleted \citep[see][for a discussion in the context of internal photoevaporation]{Cla01}. For the rate of mass loss carried in the thermal wind $\dot{M}_\mathrm{wind}$, we apply the \textsc{Fried} grid \citep{Haw18b} for a given outer disc radius $R_\mathrm{d}$, disc mass $M_\mathrm{d}$, $F$ and $m_*$. This is combined with a viscous disc evolution model, discussed below, to calculate PPD destruction time-scales.

\subsubsection{Viscous disc evolution model}

We calculate the one-dimensional viscous disc evolution using the method of \citet[][and subsequently \citealt{And13,Ros17,Win18b}]{Cla07}. In such a parametrization, viscosity is assumed to scale linearly with radius $r$ within the disc, which corresponds to a temperature profile which scales with $r^{-1/2}$ and a constant $\alpha$-viscosity parameter \citep{Sha73}. Conveniently, this prescription has similarity solutions for the disc evolution \citep{Lyn74}. For a viscosity which is proportional to $r$, such a solution for the surface density profile of the disc has the form:
\begin{equation}
\label{eq:sigma_d}
\Sigma_\mathrm{d}=\frac{M_{\mathrm{d},0}}{2\pi R_1^2 \eta} \exp \left( -\frac{\eta}{T}\right)T^{-1.5} 
\end{equation}where $M_{\mathrm{d},0}$ is the initial disc mass, $\eta \equiv r/R_1$  with $R_1$ the initial disc scaling radius, and $T=1+t/\tau_\mathrm{visc}$ for time $t$ with $\tau_\mathrm{visc}$ the viscous time-scale at $R_1$. In quantifying the FUV induced destruction time-scale $\tau_\mathrm{FUV}$, we will vary $\tau_\mathrm{visc}$ rather than the viscosity parameter $\alpha$. These two quantities can be related by the expression:
\begin{equation}
\alpha \approx 5.4\times 10^{-3} \left(\frac{\tau_\mathrm{visc}}{ \mathrm{Myr}}\right)^{-1} \left(\frac{R_1}{40\, \mathrm{au} }\right)^{3/2} \left( \frac{m_*}{M_\odot}\right)^{-1/2}.
\end{equation} Initially we truncate the surface density outside $R_{\mathrm{d},0}=2.5R_1$ to ensure a well-defined outer radius. We further assume that the initial disc mass is $M_{\mathrm{d},0} = 0.1 \,m_*$ for all of our calculations \mbox{\citep[e.g.][]{Andr13, Pas16}}. 

Numerically, the evolving surface density is defined over a one-dimensional grid evenly spaced in $r^{1/2}$ in the range $0.5$--$800$~au, with $500$ cells. A zero torque boundary condition is applied at the inner edge, and the cell at the outer edge experiences a mass flux due to both the viscous outflow and the FUV-induced wind \citep[with loss rate $\dot{M}_\mathrm{wind}$ obtained by interpolating over the \textsc{Fried} grid --][]{Haw18b}. The outer edge evolves at each timestep depending on whether there is net mass loss or accumulation. We consider a disc to be `destroyed' if $M_\mathrm{d}<10^{-5}\, M_\odot$, although our results are insensitive to this threshold since extremely low mass PPDs are quickly depleted by photoevaporation. If the disc survives for longer than $10$~Myr, we assume that it is dispersed by internal processes, such that $\tau_\mathrm{FUV} \leq 10$~Myr throughout the parameter space.

\subsubsection{Fitting formula}

\begin{figure*}
     \subfloat[\label{subfig:ff_phot_tvisc01} $\tau_\mathrm{visc} = 0.1$~Myr] {%
       \includegraphics[width=0.45\textwidth]{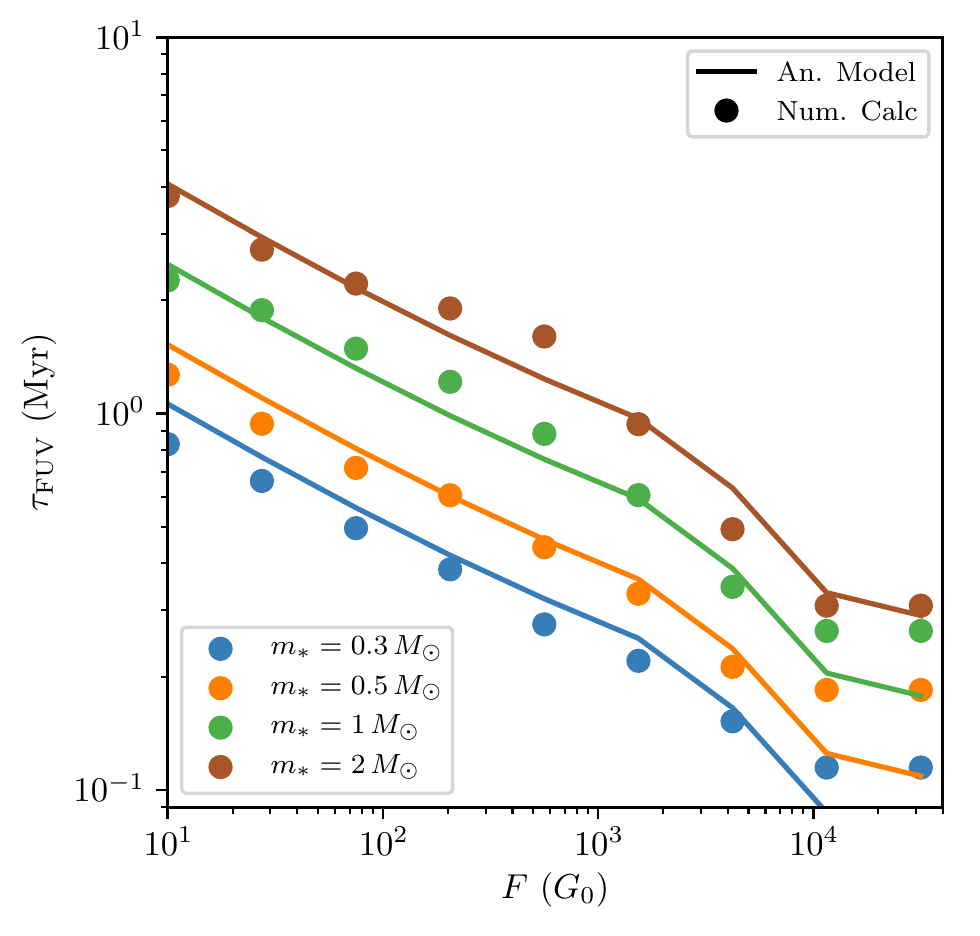}
     }  
     \subfloat[\label{subfig:ff_phot_tvisc02}$\tau_\mathrm{visc} = 0.2$~Myr ]{
       \includegraphics[width=0.45\textwidth]{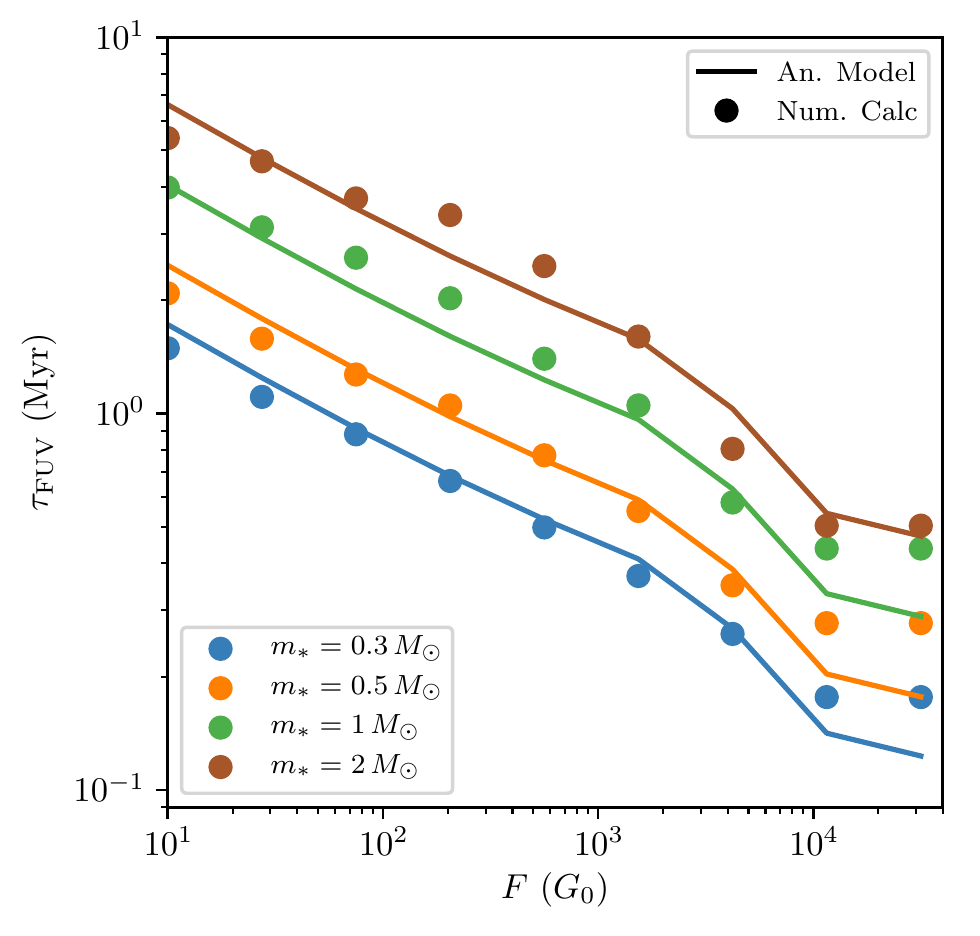}
     } \\
      \subfloat[\label{subfig:ff_phot_tvisc05}$\tau_\mathrm{visc} = 0.5$~Myr] {%
       \includegraphics[width=0.45\textwidth]{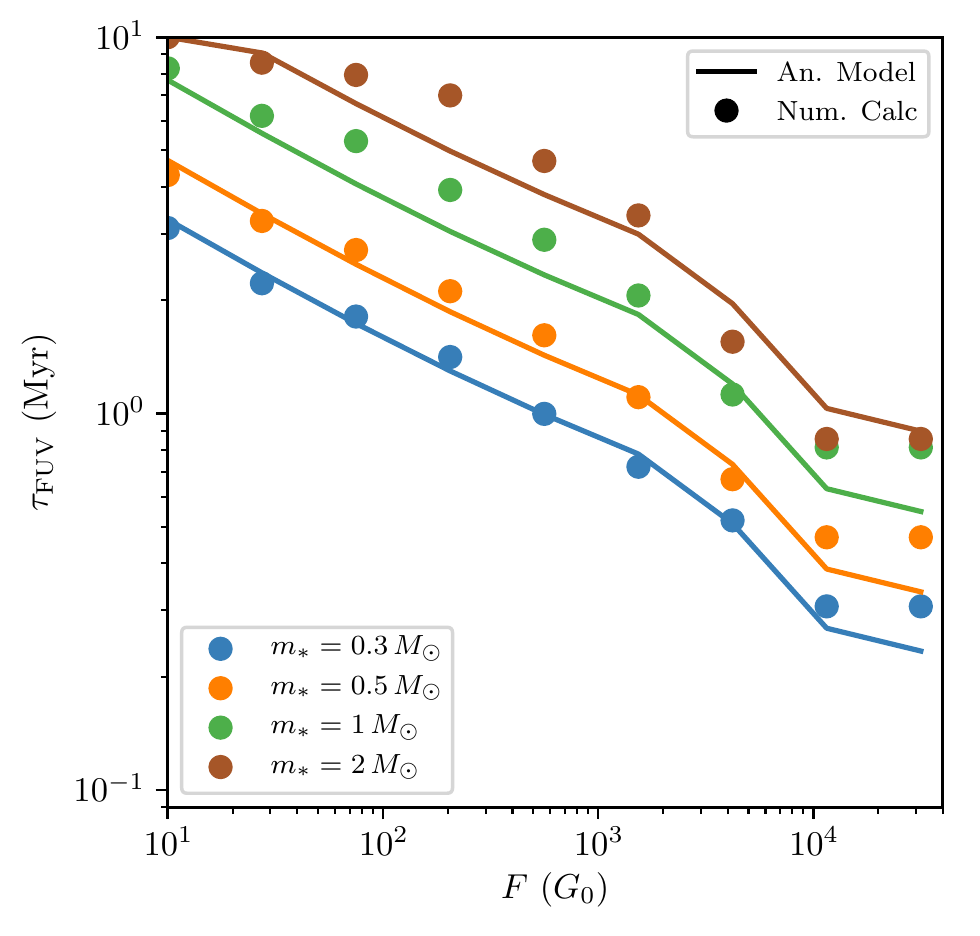} 
     }  
     \subfloat[\label{subfig:ff_phot_tvisc10} $\tau_\mathrm{visc} = 1$~Myr]{%
       \includegraphics[width=0.45\textwidth]{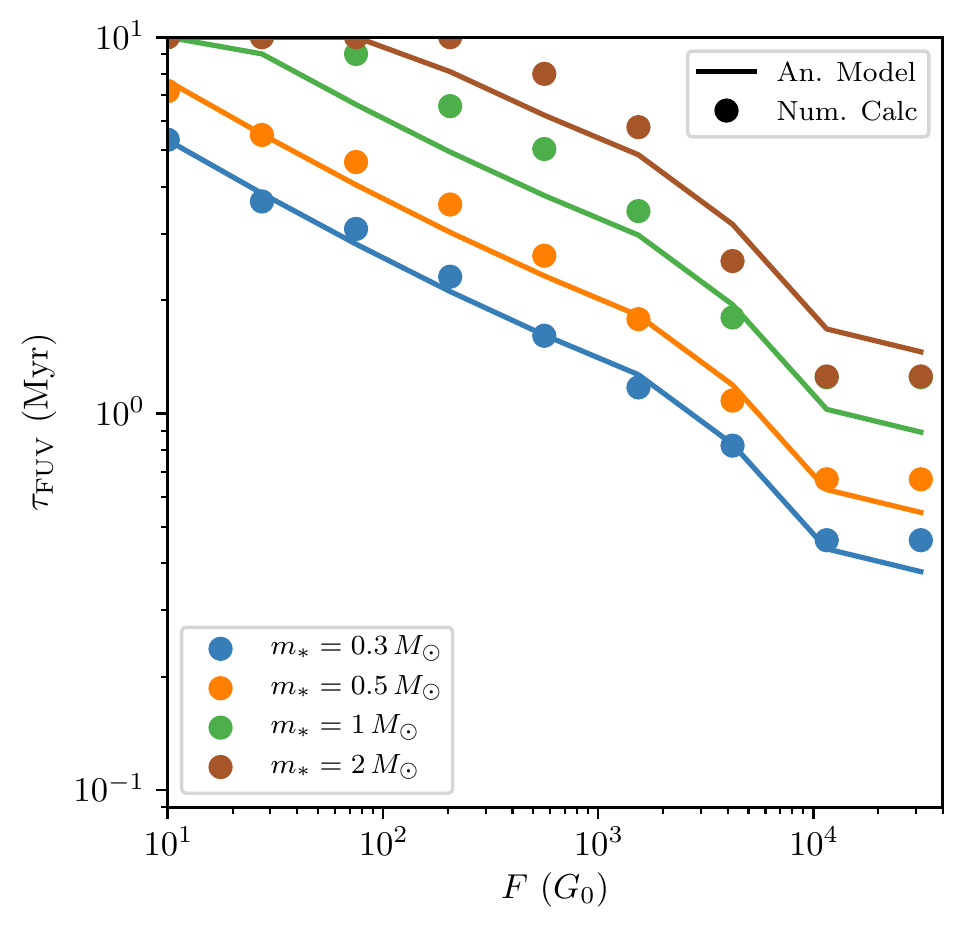}
     }
     \caption{Time-scale for disc depletion due to FUV photons, $\tau_\mathrm{FUV}$ for varying FUV flux $F$. Results are shown for different viscous time-scales $\tau_\mathrm{visc}$ and stellar host masses $m_*$. Circular markers indicate the calculation using a viscous disc evolution model, while the lines are the value using our analytic fit (equation~\ref{eq:ff_tauFUV}). }
     \label{fig:ff_phot}
   \end{figure*}

We impose a simple fitting formula to PPD FUV-induced destruction time-scale:
\begin{multline}
\label{eq:ff_tauFUV}
\tau_\mathrm{FUV} = \theta_0 \left(\frac{\tau_\mathrm{visc}}{ \mathrm{Myr}} \right)^{\theta_1}\left(\frac{m_*}{ M_\odot} \right)^{\theta_2}\left\{\exp\left[ -\left(\frac{F}{5\cdot 10^3\, G_0 } \right)^{\theta_3}\right] +1\right\} \times \\
\times \left[\left(\frac{F}{5\cdot 10^3\, G_0 }\right)^{-\theta_4}  + 1\right]\, \mathrm{Myr},
\end{multline} where $\theta_{0,1,2,3}$ are fitting parameters. In equation~\ref{eq:ff_tauFUV} we have imposed an effective minimum destruction time-scale for $F\gtrsim 10^4\, G_0$. For $F\gtrsim 10^4\, G_0$ the temperature in the photodissociation region ($\sim 10^4$~K) is insensitive to $F$ and the mass flux in the thermal wind remains approximately constant as discussed above. We have also included an intermediate regime ($10^3 \, G_0 \lesssim F \lesssim 10^4 \, G_0$) where $\tau_\mathrm{FUV}$ drops rapidly with increasing $F$; here FUV driven winds dominate mass loss throughout the lifetime of the disc. At lower $F$, the power-law relationship is weaker since accretion rates are comparable to wind driven mass loss. Fitting this formula, we obtain the values summarised in Table~\ref{table:ffpars}. In Figure~\ref{fig:ff_phot}, the numerical calculations based on the one-dimensional viscous evolution models are compared with the analytic estimate from equation~\ref{eq:ff_tauFUV}. The fitting formula reproduces the results to within a factor of order unity throughout the parameter space.


\begin{table}
\centering 
 \begin{tabular}{c c c c c c} 
 \hline
 & ${\theta_0}$  &${\theta_1}$ & ${\theta_2}$  & ${\theta_3}$ & ${\theta_4}$ \\ [0.5ex] 
 \hline
 ${\theta}_i $& 0.59  & 0.70  & 0.71 & 3.9 & 0.36 \\
 Associated var. & $\tau_\mathrm{FUV}$ & $\tau_\mathrm{visc}$ & $m_*$ & $F$  & $F$\\
 [1ex] 
 \hline
\end{tabular}
\caption{Table of fitting parameters for equation~\ref{eq:ff_tauFUV}.} 
\label{table:ffpars}
\end{table}

\subsection{Dynamical encounters}

The influence of dynamical encounters on PPD evolution has been investigated extensively \citep[e.g.][]{Ost94,Hal96, Pfa05, Olc06, Pfa06, Bre14, Win18}, and we do not expand upon the findings of those previous studies here. We instead use the result that multiple distant encounters have little effect on a disc in comparison to a single close encounter \citep{Ost94,Win18}. This means that we are free to limit our consideration to the time-scale on which one such close encounter occurs in a given environment. A consequence of this is that the tidal destruction time-scale is practically independent of the viscous evolution time-scale.

Following \citet[][see also \citealt{Ost94}]{Bin87} we can relate the impact parameter $b$ for a given encounter to the closest approach distance $z_\mathrm{min}$:
\begin{equation}
    b^2 = z_\mathrm{min}^2 \left(1+\frac{Gm_\mathrm{tot}}{v_\infty^2 z_\mathrm{min} } \right)
\end{equation} where the second term in the brackets corresponds to gravitational focusing and $v_\infty$ is the relative velocity of the two stars at infinity. The total mass $m_\mathrm{tot} = m_* + m_\mathrm{p}$ is the sum of the host and the perturber mass. Integrating over a Boltzmann distribution for $v_\infty$, we can write the differential encounter rate:
\begin{equation}
\label{eq:diffenc}
\mathrm{d} \mathcal{E} = \frac{2\sqrt{\pi} G m_\mathrm{tot} \rho_*}{\langle m_* \rangle \sigma_{v*}} \left( 1+ \frac{ 4 \sigma_{v*}^2 z_\mathrm{min} }{Gm_\mathrm{tot}} \right) \xi_*(m_\mathrm{p}) \ \mathrm{d}z_\mathrm{min} \, \mathrm{d} m_\mathrm{p},
\end{equation} where $\sigma_{v*}$ is the local 1D stellar velocity dispersion and 
\begin{equation}
\label{eq:imf}
\xi_*(m_*)\propto \begin{cases}
               m_*^{-1.3} \quad \mathrm{for } \, 0.08 \, M_\odot \leq m_*< 0.5 \, M_\odot\\ 
              m_*^{-2.3} \quad \mathrm{for } \, 0.5 \, M_\odot \leq m_* < 100 M_\odot \\
              0 \qquad \quad \, \mathrm{otherwise}
            \end{cases},
\end{equation}is the \citet{Kro01} IMF, where $\xi_*(m_*)$ is normalised and continuous. Such an IMF gives a mean stellar mass $\langle m_* \rangle \approx 0.5 \, M_\odot$. For a fixed $\sigma_{v*}$, integrating equation~\ref{eq:diffenc} over the relevant range of $z_\mathrm{min}$ gives an overall encounter rate for encounters with closest approach distances smaller than $\mathcal{Z}_\mathrm{min}$:
\begin{multline}
\label{eq:dEdmp}
    \frac{\partial \mathcal{E}}{\partial m_{\mathrm{p},0}} =\frac{0.15\rho_{*,4}\mathcal{Z}_{\mathrm{min},2}} {\sigma_{v*,0 }} \times \\ \left(m_{*,0} + m_{\mathrm{p},0} +  0.23 \mathcal{Z}_{\mathrm{min},2}\sigma_{v*,0 }^2 \right)\xi_*(m_\mathrm{p}) 
\end{multline} where 
$$
\rho_{*,4} \equiv \frac{\rho_* }{10^4\, M_\odot \, \mathrm{pc}^{-3}} ; \, \sigma_{v*,0 } \equiv \frac{\sigma_{v*} }{1\, \mathrm{km/s}} ; \, \mathcal{Z}_{\mathrm{min},2} \equiv \frac{\mathcal{Z}_\mathrm{min}}{100 \, \mathrm{au}}
$$ and $m_{*/\mathrm{p},0} \equiv {m_{*/\mathrm{p}}}/{1 \, M_\odot}$.

The problem is now reduced to finding the appropriate value for $\mathcal{Z}_\mathrm{min}$ as a function of stellar mass. For a given encounter distance, the degree to which a disc is depleted also depends on the orientation, mass ratio and eccentricity of the encounter \citep[e.g.][]{Ost94, Olc12, Bre14, Win18, Win18b}. Encounters are also more destructive if both stars host extended PPDs, where strongly interacting circumstellar material results in increased angular momentum exchange and disc mass loss \citep[e.g.][]{Pfa05b, Mun15}. However, a logical definition of a `destructive encounter' is one after which the independently evolving disc does not survive long post-encounter. We introduce the gravitational radius \citep[e.g.][]{Hol94}:
\begin{equation}
    R_g = \frac{Gm_*}{c_\mathrm{s}^2} = 8.9 \left( \frac{m_*}{1\, M_\odot} \right)\, \mathrm{au},
\end{equation}which is the radius at which photoionised gas is unbound from the stellar host ($c_\mathrm{s}\sim 10$~km/s is the sound speed in ionised gas of temperature $\sim10^4$~K). Internal photoevaporation drives thermal winds from $R_g$, resulting in a gap opening up at this radius and the quenching of viscous mass flow to the inner disc.  \citet{Cla01} demonstrate that, after this gap opens, the short viscous time-scales at small radii lead to rapid dispersal of material inwards of $R_g$. Our threshold for a destructive encounter should therefore be one which plays the role of photoevaporative dispersal in that the encounter removes the majority of mass outwards of $\sim R_g$. For equal mass, parabolic, prograde star-disc encounters, this corresponds to $\mathcal{Z}_\mathrm{min} \approx R_g/0.28$~au \citep[e.g.][]{Bre14}. The separation required to induce significant angular momentum loss during a parabolic encounter scales with $(m_\mathrm{p}/m_*)^{1/3}$ \citep{Win18}, hence we have:
\begin{equation}
    \mathcal{Z}_{\mathrm{min},2} = 0.31 \,  m_{*,0}^{2/3} m_{\mathrm{p},0}^{1/3}.
\end{equation}Taking this upper limit in the closest approach distance and integrating equation~\ref{eq:dEdmp} over $m_\mathrm{p}$ gives the encounter rate $\mathcal{E}$, or equivalently the tidal destruction time-scale:
\begin{equation}
\label{eq:tau_tidal}
    \tau_\mathrm{tidal} \equiv 1/\mathcal{E}.
\end{equation}

\begin{figure}
       \includegraphics[width=0.46\textwidth]{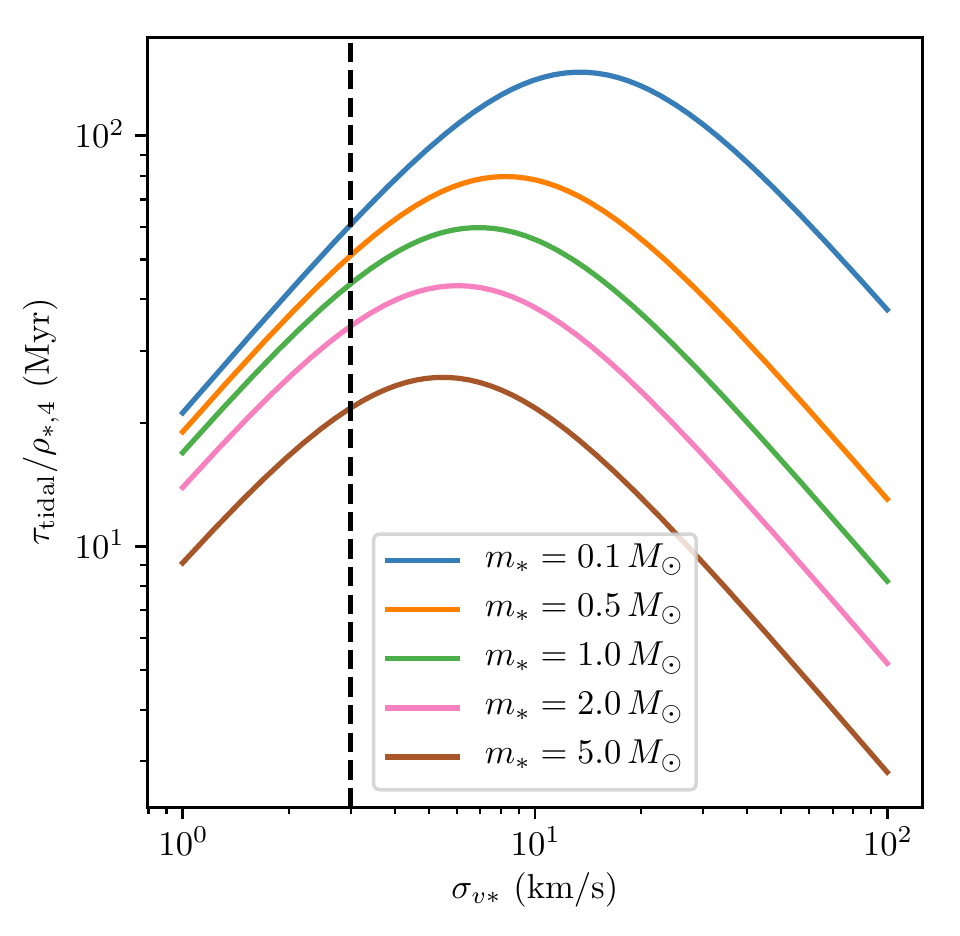}
     \caption{The encounter rate time-scale for stellar density $\rho_* = 10^4 \, M_\odot \, \mathrm{pc}^{-3}$ as a function of local velocity dispersion $\sigma_{v*}$. The dashed vertical line marks the adopted value of $\sigma_{v*}=3$~km/s in our model. }
     \label{fig:encvar}
\end{figure}

The remaining free parameter is the local stellar velocity dispersion $\sigma_{v*}$, the dependence of $\tau_\mathrm{tidal}$ on which is shown in Figure~\ref{fig:encvar}. Most star-forming regions have local $\sigma_{v*}\sim 2$--$10$~km/s, and $\tau_\mathrm{tidal}$ varies by a factor of a few in this range. We choose a fiducial $\sigma_{v*}= 3$~km/s; the resulting value of $\tau_\mathrm{tidal}$ is broadly consistent with previous findings that no significant truncation occurs for stellar number densities $\lesssim 10^4$~pc$^{-3}$  (i.e.\ $\tau_\mathrm{tidal}\gg 10$~Myr -- see \mbox{\citealt{Wij17b, Win18b}}), and that local stellar densities $> 10^5$~pc$^{-3}$ are required for dynamical encounters to act as an efficient PPD dispersal mechanism \mbox{\citep{Olc12, Vin18}}. Our choice also means that, within a reasonable range for $\sigma_{v*}$, $\tau_\mathrm{tidal}$ only varies by a factor of order a few. However, our evaluation of $\tau_\mathrm{tidal}$ should be interpreted with caution since encounters are by their nature stochastic. Our estimate is intended as a guide as to the time-scale on which severely damaging encounters occur. 

\subsection{Ram pressure stripping}

The influence of the interstellar medium is more complex to treat, since whether a PPD increases or decreases in mass is a function of both the local ISM density $\rho_\mathrm{g}$ and the ISM velocity with respect to a given host star $\vec{v}_\mathrm{g}$ \mbox{\citep{Wij17}}. Ultimately, realistic gas distributions can result in the destruction and reforming of a disc throughout the embedded phase \citep{Bat18}, and therefore discussion of time-scale for ram pressure induced disc destruction is inherently misleading. However, we can at least estimate a time-scale upon which the motion of a star through the ISM has a significant impact on the disc. This is approximately the time-scale on which the material accreted onto the disc approaches the mass of the disc itself. This can be written \mbox{\citep{Wij17}}:
\begin{equation}
\label{eq:tau_ram1}
\tau_\mathrm{ram} \sim \frac{\Sigma_\mathrm{d}}{5\rho_\mathrm{g} v_\mathrm{g}},
\end{equation}where we will assume that $v_\mathrm{g} = |\vec{v}_\mathrm{g}| \approx 1$~km/s. By choosing the initial surface density close to $R_1$, and assuming as before that $M_{\mathrm{d},0} =0.1 \, m_*$, we can use equations~\ref{eq:sigma_d} and~\ref{eq:tau_ram1} to estimate:
\begin{equation}
\label{eq:tau_ram}
\tau_\mathrm{ram} \sim \left( \frac{m_*}{M_\odot}\right) \left( \frac{\rho_\mathrm{g}}{4 \times 10^4 \, M_\odot \, \mathrm{pc}^{-3}} \right)^{-1} \, \rm{Myr}.
\end{equation} This gas density threshold ($\rho_\mathrm{g} \gtrsim 10^{-18}$~g~cm$^{-3}$) for efficient ram pressure stripping of the PPD ($\tau_\mathrm{ram}\lesssim3$~Myr) is similar to that reported in \mbox{\citet{Wij17}}. Since the response of a disc (accretion or depletion) to motion through a high gas density is uncertain, we will neglect further discussion of its influence on a PPD in this work.


\subsection{Overall dispersal time-scale}

We have now reviewed the time-scales on which three truncation processes act to deplete a PPD. Only dynamical encounters and external photoevaporation are necessarily dispersal mechanisms, and therefore when calculating the overall time for disc destruction we will focus on these two processes. For the total time-scale for PPD destruction by external influence, we therefore estimate the dispersal time-scale:
\begin{equation}
\tau_\mathrm{disp}\approx \left( \tau_\mathrm{FUV}^{-1} + \tau_\mathrm{tidal}^{-1} \right)^{-1},
\end{equation} 
where $ \tau_\mathrm{FUV}$ and $ \tau_\mathrm{tidal}$ are evaluated using equations~\ref{eq:ff_tauFUV} and~\ref{eq:tau_tidal}, respectively. We are thus able to calculate contours of constant $\tau_\mathrm{disp}$ in $\rho_*$--$F$ space. 

\section{Stellar birth environment}
\label{sec:stellar_birth}

To calculate the distribution of FUV fluxes for a stellar population, \citet{Fat08} assumed a stellar density distribution based on observed star-forming regions within $\sim 2$~kpc and extended the sample to an assumed upper limit on the number of members $N= 10^5$~stars.  Here we take a more general approach. We first relate the distribution of stellar densities to star formation physics, based on the theoretical arguments by \citet{Kru12}. This involves rewriting the lognormal gas density probability density function (PDF) in terms of the stellar overdensity in Section~\ref{sec:rhost_PDF} and relating this to the properties of the galactic disc in Section~\ref{sec:ISM_props}. We then calculate the star formation efficiency (SFE) as a function of local gas density in Section~\ref{sec:sfe}. \citet{Win18b} demonstrate that the stellar density is related to local FUV flux in the limit of high mass regions, such that we can derive the PDF of the flux $F$ from the stellar density PDF. To do this, we must also quantify the properties of neighbouring stars. The mass (and FUV luminosity) of the most massive local star decreases with decreasing mass of the star-forming region due to the stochastic sampling of the IMF, which we address in Section~\ref{sec:hostsMmax}. This is related to FUV luminosity of the most massive member in Section~\ref{sec:mvLfuv}. To obtain the fraction of stars born in a region of a certain mass, we consider the initial cluster mass function (ICMF -- Section~\ref{sec:icmf}) as a function of galactic scale gas properties. We then quantify the statistical deviation from the $\rho_*$--$F$ relationship that holds in the high-mass environment limit in Section~\ref{sec:FUVdist}, including an estimate of the FUV flux between star-forming regions. In this way, a combination of the ICMF and the stellar density PDF together determine the full, generalised 2D PDF for stellar birth environment in density--FUV space. The results of this process are presented in Section~\ref{sec:2DPDF_results} for the solar neighbourhood and the CMZ.

 Since stars form over a continuum of densities \citep{Bre10}, so far as is possible we will refrain from defining units of star formation (i.e.\ clusters/associations or GMCs). However, this definition will become necessary from Section~\ref{sec:mvLfuv}, where we address the deviation from the $\rho_*$--$F$ relationship. Some discussion of the definition of a `cluster' is required. \citet{Kru12} describes a model for the fraction of stellar clusters that remain bound, and therefore quantifies the initial bound fraction. However, we are only interested in the first few Myr of evolution; we are less concerned with whether or not a group of stars is a cluster or association. Henceforth, we will call all such groups `star-forming regions' and neglect the influence of evaporation and expansion of the stellar population.

\subsection{Stellar density PDF}
\label{sec:rhost_PDF}

Our first goal is to quantify the distribution of stellar densities as a function of galactic-scale gas properties. \citet{Kru12} framed this problem in terms of the PDF of the local gas overdensity relative to the mean density in the galactic mid-plane, $x\equiv \rho_\mathrm{g}/\rho_0$. For our purposes, it will be convenient to express the PDF in units of stellar density rather than gas density, since the former is the relevant quantity for evaluating the influence of dynamical encounters. We define $y \equiv \rho_*/\rho_0= x\epsilon(x)$, where $\epsilon$ is the local SFE. Then $y$ is the \textit{stellar} overdensity with respect to the average gas density $\rho_0$. Hence, the stellar density PDF can be written:
\begin{equation}
\label{eq:stell_PDF}
\frac{\partial p}{\partial y} \propto \frac{\partial p}{\partial x}\left(\epsilon + x\frac{\partial \epsilon}{\partial x} \right)^{-1}.
\end{equation}  However, we are in fact interested in the fraction of stars per infinitesimal region of overdensity space $\mathrm{d}y$:
\begin{equation}
\label{eq:dFdy}
\frac{\partial \mathcal{F}_*}{\partial y} \propto y \frac{\partial p}{\partial y}  \propto \frac{\partial p}{\partial x}\left(x^{-1} + \frac{\partial \ln \epsilon}{\partial x} \right)^{-1} .
\end{equation} To evaluate equation \ref{eq:dFdy}, the theoretical framework for estimating the gas density distribution and corresponding SFE is briefly reviewed below \citep[for a more complete discussion, see][and references therein]{Kru12}. 

\subsection{ISM properties}
\label{sec:ISM_props}

\subsubsection{Gas density distribution}
The PDF of the gas overdensity with respect to the mean gas density in a turbulent region is assumed to be scale-free and follows a lognormal distribution \citep[e.g.][]{Vaz94,Pad02}. It can be written as
\begin{equation}
\label{eq:overdense_PDF}
\frac{\partial p}{\partial  x}  = \frac{1}{\sqrt{2\pi \sigma^2_\rho} x} \exp \left\{ -\frac{(\ln x - \overline{\ln x})^2}{2\sigma_\rho^2} \right\},
\end{equation} where the logarithmic mean is
\begin{equation}
\label{eq:lnx}
\overline{\ln x} = - \sigma_\rho^2/2
\end{equation} and the standard deviation of the density is 
\begin{equation}
\label{eq:sig_mach}
\sigma_\rho^2 \approx \ln\left(1+ {3 b^2 \mathcal{M}^2}\right),
\end{equation}where $\mathcal{M}$ is the one-dimensional Mach number, and simulations indicate $b\approx 0.5$ \citep{Pad97, Fed10}.

Throughout this work, we will frequently refer to the properties of gas in the solar neighbourhood and in the CMZ, which we will use as regions for illustrative comparisons. The values $\mathcal{M}$ and $\rho_0$ in the galactic disc and the CMZ are discussed below in Sections~\ref{sec:gal_props}~and~\ref{sec:props}. 

\subsubsection{Connection to galactic properties}
\label{sec:gal_props}
 We now express the mid-plane density $\rho_0$ and the Mach number $\mathcal{M}$ in terms of global galactic properties. To estimate these conditions, we follow \mbox{\citet{Kru05}} in assuming the star-forming galactic disc can be modelled as a gas disc in hydrostatic equilibrium. Then we can write an expression for the \citet{Too64} $Q$ parameter in terms of the mean gas surface density $\Sigma_0$ and angular velocity $\Omega$ set by the galactic rotation curve:
 \begin{equation}
 \label{eq:Qpar}
 Q \equiv \frac{\kappa \sigma_v}{\pi G \Sigma_0} \approx \frac{\sqrt{2}\Omega \sigma_v}{\pi G \Sigma_0} ,
 \end{equation} where the epicyclic frequency $\kappa=\sqrt{2}\Omega$ for a galaxy with a flat rotation curve, and $\sigma_v$ is the one-dimensional velocity dispersion. Using equation \ref{eq:Qpar}, the mid-plane density for a disc in hydrostatic equilibrium and with scale height $h_0$ is:
\begin{equation}
\label{eq:rho0}
\rho_0 =  \frac{\Sigma_0}{2 h_0} = \frac{\pi G l_P \Sigma_0^2}{2\sigma_v^2} = \frac{l_P \Omega^2}{\pi G Q^2},
\end{equation} where $l_P\approx 3$ is a correction factor for the stellar contribution to the gravitational potential. Considering typical sound speeds in star-forming regions ($\sim 0.3$~km/s), the corresponding Mach number is approximately:
\begin{equation}
\label{eq:mach}
\mathcal{M} \approx 0.028\,   l_{\bar{P}}^{1/8}\, Q \left( \frac{\Omega}{1\, \mathrm{Myr}^{-1}}\right) ^{-1}  \frac{\Sigma_0}{1 \, M_\odot\,\mathrm{pc}^{-2}} 
\end{equation} where:
\begin{equation}
    \label{eq:lbarP}
l_{\bar{P}} \approx 10-8 f_\mathrm{GMC}
\end{equation} is the ratio of the mean pressure in a GMC to the mid-plane pressure, and $f_\mathrm{GMC}$ is the fraction of the ISM mass in GMCs. Empirically, the fraction of molecular gas $f_\mathrm{GMC}$ is related to the mean surface density \citep{Won02, Ros05b}:
\begin{equation}
\label{eq:fGMC}
f_\mathrm{GMC} \approx \left[1+2.5 \left( \frac{\Sigma_0}{10\, M_\odot \, \mathrm{pc}^{-2}}\right)^{-2}\right]^{-1}.
\end{equation} Finally, a range of values $0.5<Q<6$ are observed \citep{Ken89,Mar01}; we will explore how our results vary with $Q$.

\subsubsection{Properties of the solar neighbourhood and the CMZ}
\label{sec:props}
Throughout this work we will use the comparative examples of the solar neighbourhood and the CMZ, with parameters as follows. For the solar neighbourhood, we choose a canonical value of $Q=1.5$, $\Omega = 2.6 \times 10^{-2}$~Myr and $\Sigma_0 =12\,M_\odot$~pc$^{-2}$, in line with \citet{Kru12}. The CMZ occupies the central $\sim 250$~pc in galactocentric radius of the Milky Way, and exhibits gas properties which vary significantly from those of the disc \citep[e.g.][]{Kru13,Mol14}. The surface density in the CMZ is $\Sigma_0 \sim 1000$~$M_\odot$~pc$^{-2}$ \citep{Gue83,Hen16b}. We follow \citet{Kru14} in adopting the same \citeauthor{Too64} parameter as in the disc ($Q=1.5$) for the star-forming circumnuclear stream in the CMZ, at a radius of $\sim100$~pc. \citet{Kru15b} find an angular velocity for the stream of $\Omega \approx 1.7$~Myr$^{-1}$.

\subsection{Star formation efficiency}
\label{sec:sfe}

\subsubsection{Star formation efficiency per free-fall time}

Assuming star formation proceeds on a free-fall time $\tau_\mathrm{ff}$, the SFE $\epsilon$ can be expressed in terms of the star formation efficiency per free fall time, $\epsilon_\mathrm{ff}$. There remains debate on the exact value of $\epsilon_\mathrm{ff}$ \citep[e.g.][]{Elm02, Kru07, Elm07,Pad11,Barn17,Ler17,Hir18,Uto18,Krum19}. While in some regions (often on sub-GMC scales) the value has been found to be up to a factor $\sim 5$ higher \citep{Eva09, Hir18},  $\epsilon_\mathrm{ff} \approx 0.01$ is found across a wide dynamic range, and we will use this fiducial value in this work.

\subsubsection{Star formation time-scale}

The integrated SFE at a given density is dependent on the time for which star formation is allowed to proceed, as a multiple of the free-fall time-scale. The free-fall time-scale at local density $\rho_\mathrm{g}$ is 
\begin{equation}
\label{eq:tau_ff}
\tau_\mathrm{ff} = \sqrt{\frac{3 \pi}{32 G \rho_\mathrm{g}}},
\end{equation} and the associated SFE is
\begin{equation}
\label{eq:sfe_fb}
\epsilon_\mathrm{fb} = \frac{\epsilon_\mathrm{ff}}{\tau_\mathrm{ff}} \tau_\mathrm{fb},
\end{equation} where $\tau_\mathrm{fb}$ is the feedback time-scale, the time it takes to halt star formation. The feedback time-scale can be written as the sum of the time until the first supernova ($\tau_\mathrm{sn}\sim 3$~Myr)  plus the subsequent time until pressure equilibrium between feedback and the surrounding ISM is reached. We refer readers interested in the derivation of $\tau_\mathrm{fb}$ to \citet{Kru12} and simply quote the result of the calculation here:
\begin{equation}
\label{eq:tau_sn}
\tau_\mathrm{fb} = \frac{\tau_\mathrm{sn}}{2} \left( 1+ \sqrt{1+\frac{2\pi^2 G^2 \tau_\mathrm{ff} Q^2\Sigma_0^2}{\Phi_\mathrm{fb}\epsilon_\mathrm{ff} \tau_\mathrm{sn}^2 \Omega^2 x }}\right).
\end{equation} where $\Phi_\mathrm{fb}$ is a constant which represents the rate at which feedback injects energy into the ISM per unit stellar mass. Its exact value is uncertain \citep{Sil97, Mac99, Efs00, Aba03, Dib06}, and we use an order of magnitude estimate $\Phi_\mathrm{fb} \approx 3.2 \times 10^{32}$~erg~s$^{-1}$~$M_\odot^{-1}$ \citep[see Appendix B in][and references therein]{Kru12}.

Fundamentally, this feedback model only includes the energy deposition by supernovae, whereas observations show that `early' feedback mechanisms like photoionisation and stellar winds dominate GMC dispersal \citep[e.g.][]{Kru19,Che19}. However, the purpose of our model is not to accurately model the details of the feedback process, but to have a reasonable description of the balance between the energy output of a young stellar population and the kinetic energy density of the ambient interstellar medium. The energy in supernovae is a good proxy for the former \citep{Age13}, and \citet[Appendix C]{Kru12} show that including other feedback mechanisms does not significantly alter the population-integrated SFE when integrating over the complete density PDF.

Where star formation time-scales are long, we wish to limit our consideration to stars which host a disc, i.e.\ with ages $\lesssim 10$~Myr. In the case where the overdensity $x\rightarrow 0$, we have large $\tau_\mathrm{fb} \propto x^{-3/4}$ and $\epsilon_\mathrm{fb} \propto x^{-1/4}$. For $\tau_\mathrm{fb}>10$~Myr, we therefore limit our definition of the SFE to those stars formed within $\tau_\mathrm{inc}=10$~Myr of the onset of star formation; at these low overdensities, the `incomplete' SFE is:
\begin{equation}
\label{eq:sfe_inc}
\epsilon_\mathrm{inc} = \frac{\epsilon_\mathrm{ff}}{\tau_\mathrm{ff}} \tau_\mathrm{inc}.
\end{equation}

In general, we can write the SFE as the minimum of the feedback-limited SFE, the incomplete SFE, and the maximum local SFE. The latter limit is the SFE of protostellar cores $\epsilon_\mathrm{core}$, obtained by factoring in mass-loss through outflows. We choose $\epsilon_\mathrm{core}=0.5$, consistent with the range $0.25<\epsilon_\mathrm{core}< 0.7$ found by \citet{Mat00}. This maximum SFE is attained in the limit of high density. Hence, choosing a different value for $\epsilon_\mathrm{core}$ would just shift the PDF in stellar overdensity above some threshold ($\epsilon_\mathrm{fb}>\epsilon_\mathrm{core}$) by a factor $<2$ (taking $\partial \ln \epsilon /\partial x\rightarrow 0$ in equation~\ref{eq:dFdy}). We define the SFE as a function of overdensity:
\begin{equation}
\label{eq:SFE_min}
\epsilon =\left\{\epsilon_\mathrm{fb}^{-1}+ \epsilon_\mathrm{inc}^{-1}+\epsilon_\mathrm{core}^{-1} \right\}^{-1}.
\end{equation}We have chosen this form instead of taking the minimum of the SFEs such that $\epsilon$ is differentiable, and therefore equation~\ref{eq:dFdy}  yields a continuous PDF in density space.

\subsection{Hosts of massive stars}
\label{sec:hostsMmax}
In the remainder of this section we convert the stellar density and maximum local star mass distributions into an FUV flux spectrum for fixed local gas overdensity $x$. When quantifying the FUV flux in star-forming regions, it will be necessary to know whether or not it hosts a massive star. \citet{Fat08} showed that above a certain mass, the contribution of stars to the UV field decreases due to a flattening of the luminosity-mass function \citep[while the IMF remains steep -- see also][]{Arm00}. A different realisation of this is apparent in the study of \citet{Win18b}; they found that for regions of mass $\gtrsim 10^3 \, M_\odot$, FUV field strength is no longer strongly variable with maximum stellar mass, but is related to the local stellar density. This is a consequence of a well sampled IMF in high stellar mass environments. It is therefore necessary to delineate regions for which stellar density effectively determines FUV flux, from those that are strongly influenced by stochastic variations in the most massive local star.

For the fraction of the stellar population that are born into environments without a well-sampled IMF, calculating the FUV radiation field requires estimating the most massive stellar component, $m_\mathrm{max}$. \citet{Mas08} find that the observed distribution of $m_\mathrm{max}$ is consistent with random drawing from the IMF. We choose a \citet{Kro01} IMF (equation~\ref{eq:imf}) truncated above $100 \, M_\odot$ because this is the upper limit of the stellar atmosphere models we adopt in Section~\ref{sec:mvLfuv}. This is not a problem since the regions with stellar mass $M_\mathrm{c}<10^3 \, M_\odot$ practically always have $m_\mathrm{max} \ll 100 \, M_\odot$ (and for more massive environments, the maximum FUV luminosity is a weak function of total stellar mass).

\subsection{Environment and stellar luminosity}
\label{sec:mvLfuv}
\begin{figure}
       \includegraphics[width=0.46\textwidth]{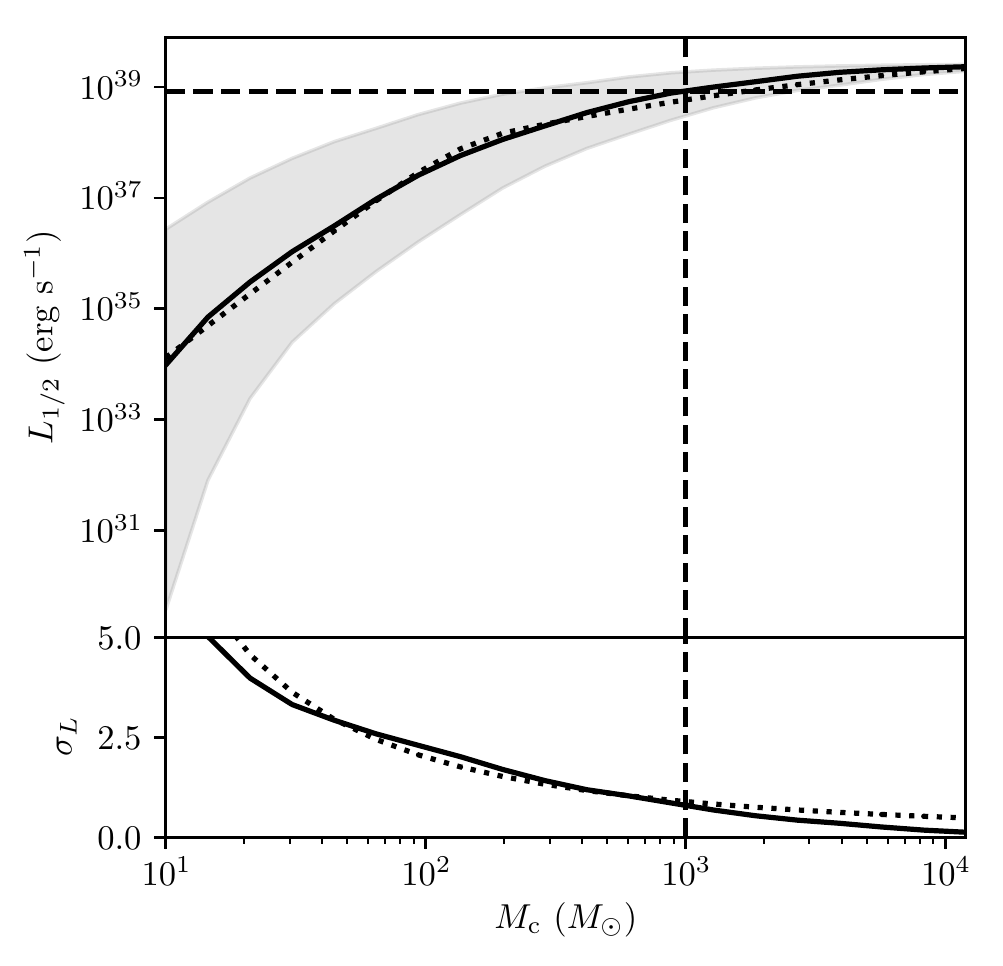}
     \caption{Top: The median luminosity $L_\mathrm{1/2}$ of the most massive star in a star-forming region with stellar mass $M_\mathrm{c}$ with the IMF described by equation~\ref{eq:imf}. The dotted line follows the analytic approximation, equation~\ref{eq:Gamma_def}. The vertical dashed line is at the critical mass $M_\mathrm{crit}\approx 10^3$~$M_\odot$ beyond which the local FUV flux is well determined by equation~\ref{eq:F0_HM}. The associated critical luminosity $L_\mathrm{crit} \approx 8.4 \times 10^{38}$~erg~s$^{-1}$~cm$^{-2}$ is shown as a horizontal dashed line. The shaded region represents the $1\, \sigma$ deviation in $L_{1/2}$. Bottom: The solid line is the logarithmic standard deviation of the luminosity $\sigma_L$, with equation~\ref{eq:sigL} indicated by the dotted line.}
     \label{fig:Lhalf}
\end{figure}


We are now required to define the units of star formation such that we are able to impose some distribution for the luminosity of the most massive \textit{local} star. We wish to evaluate the dependence of the median maximum luminosity $L_{1/2}$ as a function of cluster or association stellar mass $M_\mathrm{c}$. To calculate the FUV luminosity $L$ as a function of star mass $m_*$ we follow the method of \citet[][see also \citealt{Fat08, Win18b}]{Arm00}. We use the model grids of luminosities and effective temperatures calculated by \citet{Sch92}, taking the results for metallicity $Z=0.02$ (although the luminosity for OB stars does not change significantly in the lower metallicity results) at the output closest to $1$~Myr. We then determine the wavelength-dependent luminosity from the atmosphere models by \mbox{\citet{Cas04}}. To obtain the FUV luminosity, we integrate over the energy range for FUV photons, which is $6$~eV$< h\nu <13.6$~eV. To obtain the median luminosity as a function of total stellar mass in a region, we first draw from the IMF (equation~\ref{eq:imf}) for each $M_\mathrm{c}$ to find the median maximum star mass, $m_{1/2}$. 

Since we have chosen to assess the FUV luminosity for stars of age $1$~Myr, we must consider whether stellar evolution will significantly alter our results. In particular, we are interested in whether the FUV flux exposure of PPDs with ages $< 10$~Myr can be cut short by the death of the most massive stars. We investigate the regions of parameter space for which our static population approximation is appropriate in Appendix~\ref{sec:tOB}. We find that for the majority of star forming regions, even when the most massive star reaches the end of its lifetime, there is likely to be at least one star with comparable FUV luminosity but main sequence lifetime $\gtrsim 10$~Myr in the region. Thus, the influence of stellar evolution should not significantly effect our statistical conclusions, although may be significant in investigating disc properties in specific regions, especially for individual discs residing in close proximity to the most massive star in the region.

During the course of this work, we will regularly refer to the FUV luminosity of the most massive star in a region, and it will be useful to have an analytic approximation for this parameter. This FUV luminosity for a given total stellar mass $M_\mathrm{c}$ is shown in Figure~\ref{fig:Lhalf}. We find that for the critical mass $M_\mathrm{crit} = 10^3\, M_\odot$, we have $L_\mathrm{crit} \approx 8.3\times 10^{38}$~erg~s$^{-1}$. The results in Figure~\ref{fig:Lhalf} again justify our choice for $M_\mathrm{crit}$, since above this limit $L_{1/2}$ varies only weakly with $M_\mathrm{c}$. An analytic estimate for the median luminosity follows the form:
\begin{equation}
\label{eq:Gamma_def}
\Lambda(\phi) \equiv \frac {L_{1/2}(\phi)}{L_\mathrm{crit}} \approx \left\{ 1-e^{ - \left(f_{\mathrm{br}}\phi \right)^{\alpha} }\right\} \ln \left( 1+  \phi \right),
\end{equation} 
where we introduce 
\begin{equation}
\label{eq:phi_def}
\phi \equiv \frac{M_\mathrm{c}}{M_\mathrm{crit}} 
\end{equation} the ratio of the total stellar mass of the region to the critical mass. Equation~\ref{eq:Gamma_def} has two fitting parameters: $f_\mathrm{br} = 8.0$ and $\alpha= 2.55$. The analytic approximation in equation~\ref{eq:Gamma_def} is shown as the dotted line in the top panel of Figure~\ref{fig:Lhalf}. We further define the logarithmic deviation in the maximum luminosity:
\begin{equation}
\label{eq:sigL}
\sigma_L \approx \frac{8}{\left(3+ \log \phi \right) ^2},
\end{equation} indicated by the dotted line in the bottom panel of Figure~\ref{fig:Lhalf} (compared to the direct calculation shown as a solid line). We will further impose the limit $\sigma_L \leq 10$ for numerical reasons, although this is of little practical significance.

We emphasise that equation~\ref{eq:Gamma_def} is simply chosen as a functional form that will permit an intuition for the numerical value of physical variables and simplify our calculations in the following sections. It is appropriate in the range of $M_\mathrm{c}$ discussed here under the assumption that the maximum star mass in a region is $\lesssim 100 \, M_\odot$. 

\subsection{Initial cluster mass spectrum}
\label{sec:icmf}
In this section we are motivated to find the fraction of stars born in a star-forming region of a given mass. This is obtained via the initial cluster mass function (ICMF), which depends on the galactic environment. We follow \citet{Tru19} in assuming that the ICMF follows a modified \citet{Sch76} function, additionally truncated from below by a minimum mass:
\begin{equation}
\label{eq:icmf}
\xi_{\rm{c}} \equiv \frac{\mathrm{d}\mathcal{F}_{\rm{c}}}{\mathrm{d}M_\mathrm{c}}   \propto  \phi^{-\beta} \exp \left(- \frac{\phi_\mathrm{min}}{\phi} \right) \exp\left(-\frac{\phi}{\phi_\mathrm{max}} \right)
\end{equation} where $\beta=2$ is expected due to hierarchical collapse of molecular clouds \citep{Elm96}, and $\phi \equiv M_\mathrm{c}/10^3 M_\odot = M_\mathrm{c}/M_\mathrm{crit}$ is proportional to the stellar mass of the star forming region, as usual. Equation~\ref{eq:icmf} is then weighted by $\phi$ and normalised to give the fraction of stars born in a star-forming region of mass $\phi$. In the following we will discuss our choices for $\phi_\mathrm{min}$ and $\phi_\mathrm{max}$.

\subsubsection{Maximum cluster mass}
We follow \citet{Rei17} in calculating the maximum stellar mass in a region $\phi_\mathrm{max}$ by considering the most massive molecular cloud that can survive disruption by feedback. The ISM is stable to perturbations with a wavelength longer than the \citet{Too64} length,
\begin{equation}
\lambda_\mathrm{T}  = \frac{4\pi^2 G \Sigma_0}{\kappa^2}  = \frac{2\pi^2 G \Sigma_0}{\Omega^2} , 
\end{equation} and this is therefore the largest scale on which collapse can take place. The corresponding Toomre mass is:
\begin{equation}
\label{eq:MT}
M_\mathrm{T} =\frac{\pi \Sigma_0 \lambda_\mathrm{T}^2}{4} =   \frac{ \pi^5 G^2\Sigma_0^3}{\Omega^4}.
\end{equation}Considering the galactic plane as an infinite sheet, the 2D free fall time (collapse within the plane) of a region with radius $\lambda_\mathrm{T}/2$ \citep{Bur04}:
\begin{equation}
\label{eq:tau_ff_2D}
\tau_\mathrm{ff,2D} = \sqrt{\frac{\lambda_\mathrm{T}}{2\pi G \Sigma_0}} = \frac{\sqrt{\pi}}{\Omega}.
\end{equation}If the feedback time-scale $\tau_\mathrm{fb}<\tau_\mathrm{ff,2D}$ then  the collapsing region will be destroyed by this feedback before the conclusion of collapse, and hence the maximum mass of the GMC is given by:
\begin{equation}
M_\mathrm{GMC,max} = f_\mathrm{coll} M_\mathrm{T}
\end{equation} where 
\begin{equation}
\label{eq:fcoll}
 f_\mathrm{coll} = \mathrm{min}\left\{1, \frac{\langle \tau_\mathrm{fb}\rangle}{\tau_\mathrm{ff,2D}}\right\}^4
\end{equation} is the fraction of mass which survives collapse. We have introduced the feedback time across the entire region, which is:
\begin{equation}
\langle \tau_\mathrm{fb}\rangle \approx \tau_\mathrm{fb}(x=1).
\end{equation} To convert this into a maximum stellar mass, \citet{Rei17} multiply this by the SFE and the cluster formation efficiency. However, we are not interested here in whether or not a region is bound, and hence we only consider the SFE. We have:
\begin{equation}
\label{eq:phimax}
\phi_\mathrm{max} = \frac{\epsilon_\mathrm{eff} f_\mathrm{coll} M_\mathrm{T}}{M_\mathrm{crit}},
\end{equation} where we have defined an effective SFE in the high mass GMC limit. In line with \citet{Rei17}, we choose an effective SFE $\epsilon_\mathrm{eff} =0.1$. 

\subsubsection{Minimum cluster mass}
The minimum expected mass for a stellar cluster is more nuanced in this context, and depends on the definition we adopt for a `cluster'. As we have already discussed, in the context of this work we are not interested in whether or not a group of stars is initially `bound' in the sense that we aim to find the conditions that a star experiences early in evolution. However, we are interested in the bottom of the hierarchy for early mergers within molecular clouds. We follow \citet{Tru19} in deriving this minimum mass by considering a molecular cloud mass dependent SFE:
\begin{equation}
\label{eq:sfe_alt}
\tilde{\epsilon}_\mathrm{fb} (M_\mathrm{GMC}) = \frac{\epsilon_\mathrm{ff} }{\tilde{\tau}_\mathrm{ff}}  \tilde{\tau}_\mathrm{fb}; \qquad \tilde{\epsilon} = \mathrm{min} \left\{ \tilde{\epsilon}_\mathrm{fb},  \epsilon_\mathrm{core}\right\}.
\end{equation}  We find that $\tilde{\epsilon}_\mathrm{fb}$ increases with decreasing $M_\mathrm{GMC}$ for small $M_\mathrm{GMC}$, such that below a certain cloud mass:
\begin{equation}
\tilde{\epsilon} \gtrsim {\epsilon}_\mathrm{th} \approx 0.2,
\end{equation} where the threshold SFE $\epsilon_\mathrm{th}$ is given by the efficiency required to produce a bound cluster after instantaneous gas expulsion \citep{Bau07}. For cloud masses below this limit $M_\mathrm{th}$, SFE is high enough to result in hierarchical merging into single objects (which can be considered to be associated within our context), and the minimum mass for a star forming region can be written:
\begin{equation}
\phi_\mathrm{min}{M_\mathrm{crit}} = {\epsilon_\mathrm{th} M_\mathrm{th}}.
\end{equation}We are now left with the problem of solving the equations for the SFE with respect to the galactic scale ISM properties. 

In the numerical derivation of $\phi_\mathrm{min}$ we consider the SFE across an entire molecular cloud, $\tilde{\epsilon}$, as opposed to the local SFE considered in Section~\ref{sec:sfe}, ${\epsilon}$, which is dependent on $x$ and consistent with the calculation of \citet{Kru12}. The primary difference is that in the former case, we can estimate the supernova time-scale $\tilde{\tau}_\mathrm{sn}$ based on the local stellar mass (but not the influence of density), while in the latter we can assess the influence of local density on the feedback efficiency (but not the variation in supernova time-scale). Ideally we would consider the SFE as a function of both cloud mass and local density. However, this would greatly complicate our prescription, in which we need to define the flux PDF at each stellar density. Instead, we are content to consider $\tilde{\epsilon}$ for the purposes of assessing the minimum stellar mass in a region since these two different prescriptions are physically compatible; $\tilde{\epsilon} $ being SFE on a GMC scale, and $\epsilon$ being SFE on a local (stellar) scale.  

We refer the reader interested in the derivation of the feedback time-scale  to \citet{Tru19}. In brief, the local gas density used when deriving the SFE as a function of local gas overdensity is replaced by the average cloud density:
\begin{equation}
\label{eq:rhoGMC}
\rho_\mathrm{GMC}  = \frac 3 4 \left(\frac{\pi \Sigma_\mathrm{GMC}^3}{M_\mathrm{GMC}} \right)^{1/2} =  \frac 3 4 \left(\frac{\pi \Sigma_0^3 f_\mathrm{\Sigma}^3}{M_\mathrm{GMC}} \right)^{1/2},
\end{equation}where we define $f_\Sigma \equiv \Sigma_\mathrm{GMC}/\Sigma_0$, the ratio between the GMC surface density and the mean gas surface density. Following \citet{Kru05} and \citet{Kru15}, for a virial ratio $\alpha_\mathrm{vir}=1.3$ \citep[appropriate for pressure confined GMCs --][]{Ber92} this ratio can be written:
\begin{equation}
f_\Sigma = 3.92 \left(\frac{l_{\bar{P}}}{2} \right)^{1/2}.
\end{equation}
In the solar neighbourhood, this yields $\Sigma_\mathrm{GMC}  \approx 90 \, M_\odot$~pc$^{-2}$ \citep[consistent with the findings of][]{Bol08}. The GMC mass dependent free fall time-scale can be written:
\begin{equation}
\tilde{\tau}_\mathrm{ff} = \sqrt{\frac{\pi^{1/2}}{8G}} \left( \frac { M_\mathrm{GMC}}{f_\Sigma^3 \Sigma_0^3}\right)^{1/4}.
\end{equation}Finally, we estimate the time-scale for a supernova to occur by considering the progenitor formation time-scale, which is important at low cloud masses, such that we have:
\begin{equation}
\tilde{\tau}_\mathrm{sn} = \tau_\mathrm{sn} + \Delta \tau_\mathrm{sn},
\end{equation}where we have defined $\Delta \tau_\mathrm{sn}$ as the time it takes for the stellar component of a star-forming region to reach a sufficient mass to form an OB star. This mass is calculated by \citet{Tru19} to be $M_\mathrm{OB} \approx 99  \, M_\odot$, and the corresponding time-scale:
\begin{equation}
\label{eq:tauGMC_sn}
\Delta \tau_\mathrm{sn} = \frac{M_\mathrm{OB} \tilde{\tau}_\mathrm{ff}}{M_\mathrm{GMC} \epsilon_\mathrm{ff}}.
\end{equation} With these adjustments, an alternate version of equation~\ref{eq:tau_sn} is:
\begin{equation}
\label{eq:tauGMC_fb}
\tilde{\tau}_\mathrm{fb} \approx  \frac{\tilde{\tau}_\mathrm{sn}}{2} \left[ 1+ \sqrt{1+ \frac{2\sqrt{2G} \pi^{3/4} l_P \Sigma_0^2 M_\mathrm{GMC}^{3/4} }{3\Phi_\mathrm{fb}\epsilon_\mathrm{ff} \tilde{\tau}_\mathrm{sn}^2 \Sigma_\mathrm{GMC}^{9/4}}} \right].
\end{equation}

We can solve the system of equations~\ref{eq:sfe_alt} to~\ref{eq:tauGMC_fb} for $M_\mathrm{th}$ such that
\begin{equation}
\left.\tilde{\epsilon}\right|^{M_\mathrm{th}} = \epsilon_\mathrm{th},
\end{equation}to find $\phi_\mathrm{min}$ (i.e. the minimum mass of a star forming region). From the above formulation there is no physical reason why we cannot have $\phi_\mathrm{min}>\phi_\mathrm{max}$. In this case, the bottom of the hierarchy exceeds the maximum mass that can be produced in such an environment, and the former is therefore set by the latter. This results in a narrow distribution of stellar masses, and $\phi_\mathrm{min}=\phi_\mathrm{max}$ (set by the maximum possible mass), such that our ICMF continues to be physically valid. For numerical reasons, it will also be convenient to set limits on the allowed values for $\phi_\mathrm{min}$ and $\phi_\mathrm{max}$. We define $\phi_\mathrm{max,min}=\phi_\mathrm{min,min}=10^{-2}$, $\phi_\mathrm{min,max}=100$ and $\phi_\mathrm{max, max}=10^6$.  Our results are not strongly sensitive to these choices since the FUV flux experienced by PPDs is insensitive to the stellar mass of the star-forming region in the high and low mass limits (see Section~\ref{sec:FUVdist}). 
 

\subsubsection{Derived initial cluster mass function}
\begin{figure}
    \includegraphics[width=0.46\textwidth]{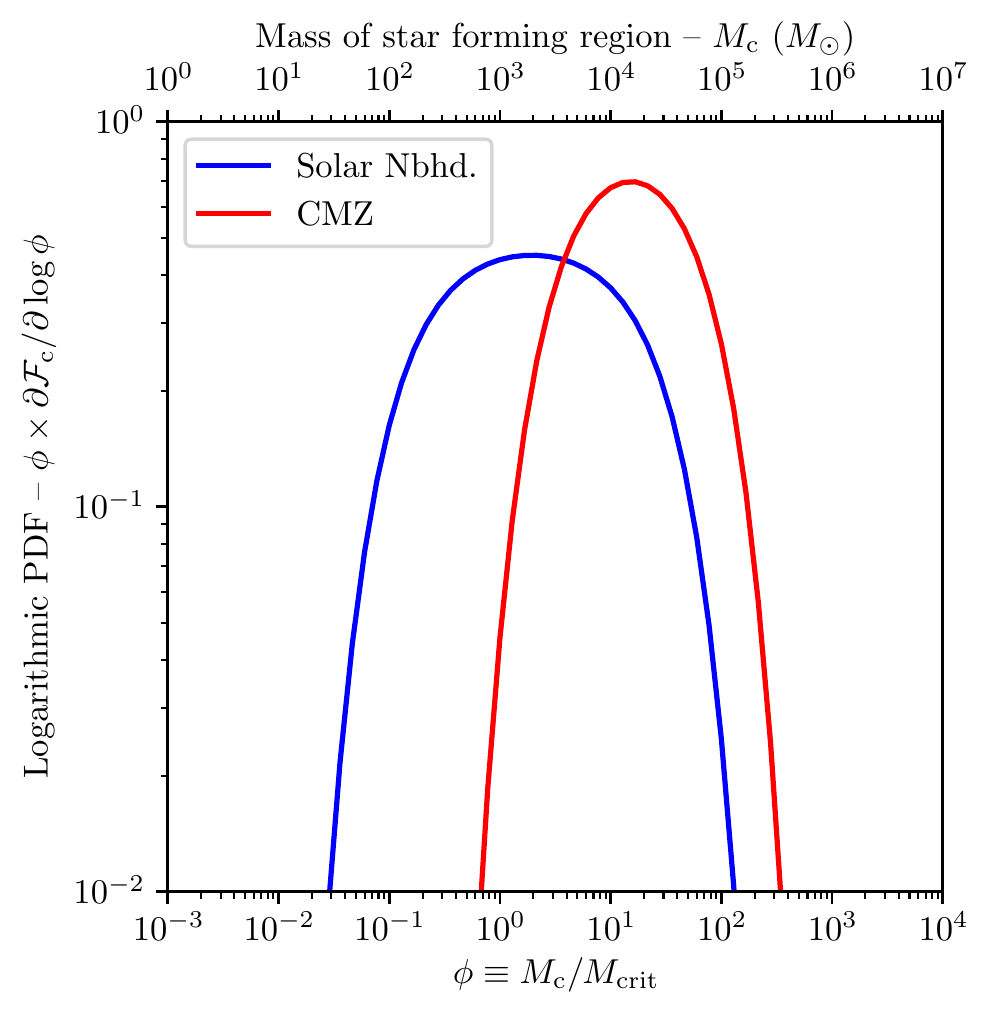}
    \caption{The ICMF in terms of $\phi\equiv M_\mathrm{c}/M_\mathrm{crit}$ weighted by the stellar mass of the region, indicating the fraction of stars born in such an environment. The lower limit $\phi_\mathrm{min}$ is given by the bottom of the single-object merger hierarchy calculated by \citet{Tru19}. The maximum stellar mass of a region $\phi_\mathrm{max}$ is the stellar component of a GMC with mass given by the feedback-limited fraction of the Toomre mass \citep{Rei17}. The blue line is for the solar neighbourhood, while the red line describes the ICMF in the CMZ. }
     \label{fig:dFdphi}
\end{figure}

The theoretical ICMFs of the solar neighbourhood and CMZ are shown in Figure~\ref{fig:dFdphi}, weighted by mass to illustrate the fraction of stars initially found in a region of a given mass. We note that our upper mass estimates are somewhat larger than those of \citet{Rei17} since we are not interested in the cluster formation efficiency. We find that regions in the CMZ
have $\phi_\mathrm{min} = 3.1$ (i.e. minimum mass $M_\mathrm{min}=3.1\times10^3 \, M_\odot$) and $\phi_\mathrm{max} = 74$ (i.e. maximum mass $M_\mathrm{max}=7.4 \times 10^4\, M_\odot$), while the solar neighbourhood has $\phi_\mathrm{min}=0.13$ (i.e. $M_\mathrm{min}=130\, M_\odot$) and $\phi_\mathrm{max}= 33$ (i.e. $M_\mathrm{max}=3.3 \times 10^4 \, M_\odot$). These adopted minimum masses are the same as those quoted in \citet{Tru19}. While we do not compare the ICMF here to the observed distribution of young star forming regions, this exercise is performed in the latter study, wherein the theoretical ICMF is found to be in good agreement with the existing observational constraints.

\subsection{FUV flux distribution}
\label{sec:FUVdist}

To build a distribution of FUV flux as a function of local density, we are motivated to quantify the expected (mean) flux $F_0$. We now outline a model motivated by theory and observations to find $F_0(x, \phi)$ for fixed ISM properties. 

\subsubsection{High mass clustered environment regime}
\label{sec:HM_psi_PDF}
For high mass star-forming regions, the FUV flux is closely related to the stellar density $\rho_*$ \citep{Win18b}. Empirically, the mean FUV flux in high mass environments is
\begin{equation}
\label{eq:F0_HM}
 F_0^\mathrm{HM}  \approx 1000 \left( \frac{\rho_*}{1\, M_\odot \, \mathrm{pc}^{-3}}\right)^{1/2}\, G_0.
\end{equation}We will assume that for small $\phi$, as the mass of the local environment increases the flux distribution approaches this average. This is an empirical relationship. Since the stars which dominate the local FUV flux \citep[of mass $\sim 30$--$50\,M_\odot$ -- e.g.][]{Arm00} make up only a small fraction of the IMF ($\sim 10^{-3}$), equation~\ref{eq:F0_HM} is determined by the radial stellar density profile of star-forming regions rather than the local density of OB stars. 

 \subsubsection{Flux in the field}
 \label{sec:field_flux}
 To define the full PDF of FUV flux for a given density, it will be further necessary to define a minimum value for which the FUV exposure is set by the field strength between star-forming regions, dependent on their separation $\lambda_0$. For an ISM which is shaped by expanding bubbles driven by stellar feedback, the separation is set by the scale on which the bubbles depressurise, which is the scale height of the disc \citep[][]{McK77, Hop12}. \citet{Kru19} recently confirmed this empirically for the nearby spiral galaxy NGC300 across all galactocentric radii in the range 0--3~kpc (or out to $\sim0.5R_{25}$). We must also consider the limit where the mean GMC radius:
 \begin{equation}
     \langle R_\mathrm{GMC}\rangle = \int \mathrm{d} \phi \, \xi_\mathrm{c} \sqrt{\frac{M_\mathrm{GMC}(\phi)}{\pi \Sigma_0 f_\Sigma} }  
 \end{equation} becomes greater than $h_0$ (regions of large $\Omega$). In this case, $M_\mathrm{GMC}(\phi)$ can be found by solving:
 \begin{equation}
     \tilde{\epsilon}(M_\mathrm{GMC}) M_\mathrm{GMC} = M_\mathrm{crit} \phi
 \end{equation} to give $\langle R_\mathrm{GMC} \rangle$. Then we have:
 \begin{equation}
     \lambda_0 = 2\cdot \max \left\{ h_0, \langle R_\mathrm{GMC}\rangle \right\}.
 \end{equation}
 
 We must also consider the extinction of FUV photons due to the surface density of gas between star-forming regions:
 \begin{equation}
     \Sigma_\mathrm{eff}^\mathrm{f} = 2\rho_0 \lambda_0 x = \frac{\Sigma_0 \lambda_0}{h_0} x,
 \end{equation}where $\rho_0$ is related to $\Sigma_0$ and $h_0$ by equation~\ref{eq:rho0}. We define an extinction factor:
\begin{equation}
C_\mathrm{ext} \equiv \frac{\Sigma_0 }{13.36 \, M_\odot \, \mathrm{pc}^{-2}}, 
\end{equation} where we have normalised the mean surface density by the column density required for  1~mag of extinction in the FUV. This normalisation is calculated from the ratio of extinction in FUV to the visible $A_\mathrm{FUV}/A_\mathrm{V} \approx 2.7$ \citep{Car89} and the column density of hydrogen required for 1~mag of extinction in the visible $N_\mathrm{H}/A_\mathrm{V} = 1.8 \times 10^{21}$~cm$^{-2}$~mag$^{-1}$ \citep{Pre95}. 

Finally, we must also consider star-forming regions occupied by many OB stars, which matters when the fractional variation of distances between sources becomes small (i.e. for a star well outside of a star-forming region). This can be accounted for weighting flux contributions by stellar mass for regions with more than one strong FUV source:
\begin{equation}
    \phi_{>1} = \max\{1, \phi\}.
\end{equation}This consideration highlights the importance of our normalisation for $\phi$, chosen such that the FUV luminosity of the most massive star is only logarithmically dependent on $\phi$ for $\phi>1$ (equation~\ref{eq:Gamma_def}). The factor $\phi_{>1}$ addresses the weighted contribution of the the most massive star-forming regions to the average FUV field in a given galactic environment.

With the above considerations, the FUV field strength between star-forming regions can now be calculated by the weighted contribution from the star-forming regions multiplied by an extinction factor:
 \begin{equation}
     F_0^\mathrm{f} =\frac{L_\mathrm{crit}}{ \lambda_0^2} \int \mathrm{d} \phi \, \phi_{>1} \xi_\mathrm{c} (\phi) \Lambda (\phi) \cdot \int \! \mathrm{d} x \, \exp\left(-\frac{ C_\mathrm{ext}\lambda_0 }{h_0}x\right)\frac{\partial p}{\partial x} .
 \end{equation}In the Solar neighbourhood this calculation yields $F_0^\mathrm{f}= 0.8\,G_0$ \citep[close to the empirical estimate by][]{Hab68} and for the CMZ we obtain $F_0^\mathrm{f} =2200\,G_0$.

 \subsubsection{Low mass clustered environments}
 \label{sec:LM_psi_PDF}
Low mass environments do not have a well sampled IMF, but may still represent regions of high density. For such a region the flux is dependent on the most massive stellar component. In a statistical sense, this is in turn dependent on the mass of the star-forming region. We require a functional form for which the average FUV flux $F_0$ at a given stellar density is proportional to the average luminosity of the most massive neighbour $\Lambda$, but is limited in the low mass limit by $F_0^\mathrm{f}$:
\begin{equation}
\label{eq:flux_general}
\psi_0 =  \Lambda + \psi_0^\mathrm{f}.
\end{equation} We have defined the ratio of the average local flux to the high mass limit $\psi_0 \equiv F_0/F_0^\mathrm{HM}$, with $\psi_0^\mathrm{f} =  F_0^\mathrm{f}/F_0^\mathrm{HM}$. The normalisation scale $F_0^\mathrm{HM}(\rho_*)$ is a function of density, and since $F_0^\mathrm{HM}\rightarrow 0$ as $\rho_* \rightarrow 0$ we have $\psi_0^\mathrm{f}\gg 1$. In this limit it follows that $\psi_0\approx \psi_0^\mathrm{f}$, and the PDF for $\psi_0$ is:
\begin{equation}
\label{eq:dpdpsi0_psi0f}
\left. \frac{\partial \mathcal{F}_*}{\partial  \psi_0}\right|^{\rho_*\rightarrow 0} \rightarrow \delta (\psi_0 -\psi_0^\mathrm{f})
\end{equation} where $\delta$ is a Dirac delta function (for fixed $\rho_*$ or, equivalently, $x$). In the lower limit ($\psi_0<\psi_0^\mathrm{f}$), no corresponding $\Lambda$ exists and we have:
\begin{equation}
\label{eq:dpdpsi0_upper}
\left. \frac{\partial \mathcal{F}_*}{\partial  \psi_0}\right|^{\psi_0< \psi_0^\mathrm{f}} = 0,
\end{equation} for all $x$. Above this threshold, we can evaluate $\Lambda$ for a given value $\psi_0$ and write the PDF:
\begin{equation}
\label{eq:psi0_PDF}
\left. \frac{\partial \mathcal{F}_*}{\partial  \psi_0} \right|^{\psi_0>\psi_0^\mathrm{f}}= \frac{\partial \mathcal{F}_*}{\partial\phi} \left|\frac{\partial \phi}{\partial \psi_0}\right| =\frac{\partial \mathcal{F}_*}{\partial\phi} \left|\frac{\partial \Lambda}{\partial \psi_0} \frac{\partial \phi}{\partial \Lambda} \right|= \frac{\partial \mathcal{F}_*}{\partial\phi}  \left|\frac{\partial \Lambda}{\partial \phi}\right|^{-1} , 
\end{equation} where $\partial \Lambda/\partial \phi$ can be obtained from equation~\ref{eq:Gamma_def}.
 
 To evaluate the PDF with respect to $\phi$, we consider the (normalised) ICMF defined in Section~\ref{sec:icmf}:
\begin{equation}
\label{eq:MCMF}
\frac{\partial \mathcal{F}_*}{\partial\phi} \propto \phi \xi_\mathrm{c}(\phi) .
\end{equation} We have multiplied the ICMF by a factor $\phi$ since the number of stars within a star-forming region scales with stellar mass. Hence the PDF for FUV flux, equation~\ref{eq:psi0_PDF}, can be expressed analytically at a fixed overdensity $x$.

We are additionally interested in the influence of extinction of FUV photons due to the molecular gas present in the nascent cluster or association. The calculation of an equivalent extincted normalised mean flux $\psi_0^\mathrm{ext}$ requires further assumptions regarding the initial distribution of stars and gas. These are reviewed in Appendix~\ref{sec:FUVext} where we calculate the quantities relevant in producing an upper limit on the influence of extinction.

\subsubsection{Dispersion from mean FUV flux}

Equation~\ref{eq:psi0_PDF} defines a PDF for the mean flux distribution for fixed density, but in deriving it we have assumed that all star-forming regions of a fixed mass exhibit the same flux distribution. This is clearly not the case, as the most massive star and the internal density profile can yield variations in the flux experienced by the stellar population. To model these variations, we consider deviations from the average flux ratio $\psi_0$ which follow a lognormal distribution:
\begin{equation}
\label{eq:lognorm_dpsi_PDF}
 \frac{\partial \mathcal{F}_*}{\partial  \delta \psi}  = \frac{1}{\sqrt{2\pi \sigma^2_F} \delta \psi} \exp \left\{ -\frac{(\ln \delta \psi)^2}{2\sigma_F^2} \right\}
\end{equation} where $\delta\psi \equiv \psi/\psi_0$ and $\psi = F/F_0^\mathrm{HM}$. The logarithmic flux dispersion $\sigma_F$ is the contribution of the dispersion $\sigma_F^\mathrm{f}$ in flux arising from varying spatial separations from ionising sources, and the dispersion $\sigma_L$ in the luminosity of the most massive member of the region. The former dominates the dispersion in the limit where FUV flux is determined by the field value, and in the limit of massive environments where $\sigma_L$ is small. In the intermediate regime, the dispersion is dominated by $\sigma_L$. Hence we have:
\begin{equation}
\label{eq:sigF}
\sigma_F = \sigma_F^\mathrm{f} +  \sigma_L (\phi) \cdot \mathcal{W} (\psi_0)
\end{equation} where we estimate $\sigma_F^\mathrm{f} =0.5$ \citep[approximated from the results of][]{Win18b}. The weighting function is defined:
\begin{equation}
\mathcal{W} =\mathrm{max} \left\{\mathrm{erf}\left(\frac{\ln \psi_0- \ln \psi_0^\mathrm{f}}{\sqrt{2}\sigma_F^\mathrm{f}}\right), 0  \right\}.
\end{equation}We have used $\sigma_F^\mathrm{f}$ as the deviation in the (logarithmic) error function such that $\sigma_{F} \rightarrow \sigma_F^\mathrm{f}$ in this range around $\psi_0^\mathrm{f}$. Otherwise the contribution from $\sigma_{L}$ could result in a significant fraction of stars falling below the field flux threshold.

Since $\psi$ is the product of $\psi_0$ and $\delta \psi$, we can evaluate its PDF using equations~\ref{eq:psi0_PDF} and~\ref{eq:lognorm_dpsi_PDF}:
\begin{equation}
\label{eq:dFdpsi}
 \frac{\partial \mathcal{F}_*}{\partial \psi} = \int \! \mathrm{d} \psi_0  \,\frac{\partial \mathcal{F}_*}{\partial  \psi_0} \frac{\partial \mathcal{F}_*}{\partial \delta \psi} \frac{1} {\psi_0}.
\end{equation} Hence, we have a PDF for the FUV flux experienced by a stellar population at a fixed overdensity $x$. 

To calculate the extincted flux, the above prescription cannot be applied, because we already needed to marginalise over $\phi$ in the initial calculation of the PDF for $\psi_0^\mathrm{ext}$ (see Appendix~\ref{sec:FUVext}). To simplify, we assume $\sigma_F = \sigma^\mathrm{f}_F$ as a first order estimate. While this underestimates the dispersion in flux for intermediate $\phi$ values, we perform these calculations to give a sense of the severity of FUV extinction for the most extreme regions. In this case, the flux dispersion is $\sigma_F \sim \sigma^\mathrm{f}_F$ anyway. As discussed in Appendix~\ref{sec:FUVext}, more detailed estimates of the true influence of extinction are required, and we leave this for future work. 

\subsection{Stellar density--FUV flux distribution}
\label{sec:2DPDF_results}

\subsubsection{No extinction}

\begin{figure*}
      \includegraphics[width=\textwidth]{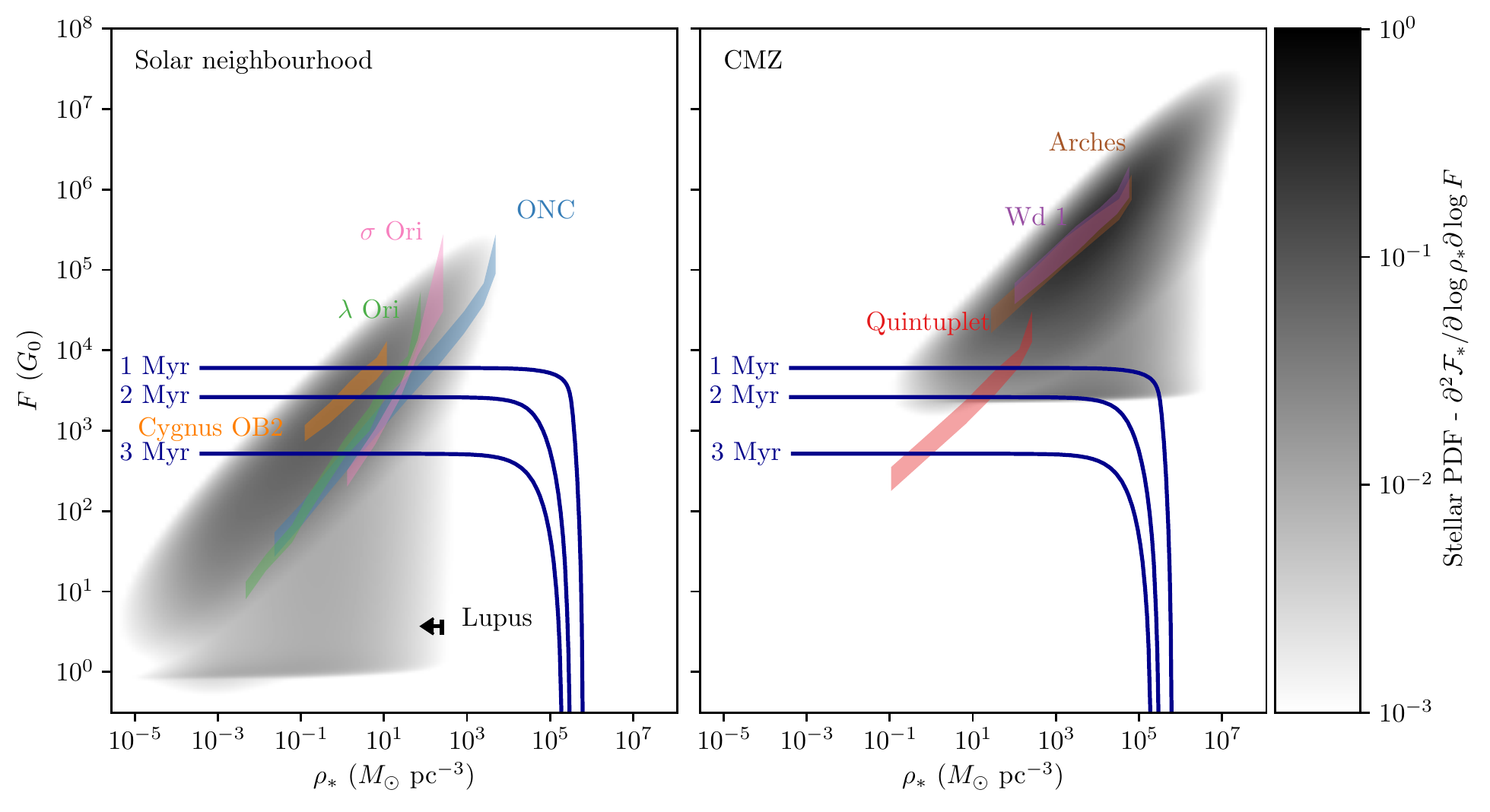}
     \caption{Two dimensional PDF for stars in $F$--$\rho_*$ (FUV flux--stellar density) space. The left panel is for the solar neighbourhood, described by mean surface density $\Sigma_0 =12 \, M_\odot$~pc$^{-2}$, \citeauthor{Too64} $Q=1.5$, and angular speed $\Omega = 2.6 \times 10^{-2}$~Myr$^{-1}$. The righ panel reflects conditions in the CMZ, with $\Sigma_0 = 1000 \, M_\odot$, $Q=1.5$ and $\Omega = 1.7$~Myr$^{-1}$. We have marked contours in the PPD dispersal time-scale calculated with the model described in Section~\ref{sec:ppd_dest} for a star of mass $m_*=0.5$~$M_\odot$ (approximately the mean mass stellar mass from our IMF) with a viscosity parameter $\alpha= 5.4 \times 10^{-3}$. We have additionally indicated some empirically derived contours calculated by \citet{Win18b} for a number of young stellar environments, truncated at a radius such that $90\%$ of stars for each region are included. }
     \label{fig:PDF_Frho}
   \end{figure*}

   To illustrate the consequences of the formulation we have presented in this section, we now apply our results to the solar neighbourhood and the CMZ with parameters indicated in Sections~\ref{sec:props}. The PDF for stars in terms of the local stellar density and FUV flux is given by:
   \begin{equation}
   \frac{\partial^2 \mathcal{F}_*}{\partial \rho_*\partial F} = \frac{\partial  \mathcal{F}_*}{\partial \rho_*} \frac{\partial  \mathcal{F}_*}{\partial F} \propto \frac{\partial  \mathcal{F}_*}{\partial y}  \frac{\partial  \mathcal{F}_*}{\partial \psi},
   \end{equation}where the last expression is evaluated using equations~\ref{eq:dFdy} and \ref{eq:dFdpsi}. The results of this calculation are shown in Figure~\ref{fig:PDF_Frho} in the case of no interstellar extinction. We have indicated contours of equal dispersal time-scale for $\tau_\mathrm{disp}=1$, $2$ and $3$~Myr for a star of mass $0.5\,M_\odot$ hosting a PPD with $\alpha=5.4 \times 10^{-3}$ (as calculated in Section~\ref{sec:ppd_dest}). 
  
  Although the sample of young star-forming regions compiled by \mbox{\citet{Win18b}} is not complete, we can qualitively compare our results in Figure~\ref{fig:PDF_Frho}, where we overplot contours for some observed star-forming environments. In agreement with \citet{Win18b}, we find that stars do not occupy regions of high density and low FUV flux such that disc dispersal would be driven by dynamical encounters (i.e. external photoevaporation dominates). In the solar neighbourhood the most extreme $F$ and $\rho_*$ lies at $F\sim 10^5$~$G_0$ and $\rho_*\sim 10^4 \, M_\odot$~pc$^{-3}$. This is equivalent to the conditions within the core of the Orion Nebula Cluster (ONC); the most extreme observed environment in the solar neighbourhood in terms of these parameters \citep[see][for a discussion of photoevaporated PPDs in such an environment]{Win19b}. The lower limit in FUV flux is $\sim 1\,  G_0$, which is the observed field value in the solar neighbourhood \citep{Hab68}. Additionally, we predict a number of regions with low $F \sim 1$--$10\, G_0$, but $\rho_* \sim 10^3 \, M_\odot$~pc$^{-3}$. This reflects the conditions observed in Lupus for example \citep{Nak00,Mer08, Cle16, Haw17}. In summary, the distribution of stellar environments is in good agreement with what we would expect from observations of local regions. 
   
  In the case of the CMZ, we find much higher typical FUV field strengths and densities. The most extreme regions lie at $\rho_* \sim 10^6 \, M_\odot$~pc$^{-3}$ and $F\sim 10^6 \, G_0$.  This is comparable to the conditions found in core of Arches and Westerlund 1 \citep{Fig99, Men07, Win18b}. The contour for Quintuplet is lower density and experiences lower FUV flux than the majority of stars as predicted by our model. This may be due to dynamical evolution of the cluster, which is older and lower mass than Arches. The velocity dispersion in Arches is $\sim 5.4$~km/s \citep{Cla12}, while Quintuplet may have a velocity dispersion as high as $10$~km/s \citep{Sto14}. Given Quintuplet's present day stellar density, this upper limit is consistent with a supervirial dynamical state, such that it is possible that it has undergone an epoch of expansion. In addition, the contours presented in \citet{Win18b} used a conservative estimate for the maximimum stellar mass, and did not account for the contribution of the field flux at large radii. In general, the distribution of stellar birth environments in the CMZ suggest that both FUV photons and dynamical encounters play a role in PPD evolution (although for the majority of discs, external photoevaporation remains the dominant dispersal mechanism), and that discs cannot survive for long in such environments. 
   
\subsubsection{Maximal Extinction}

  \begin{figure*}
    \includegraphics[width=\textwidth]{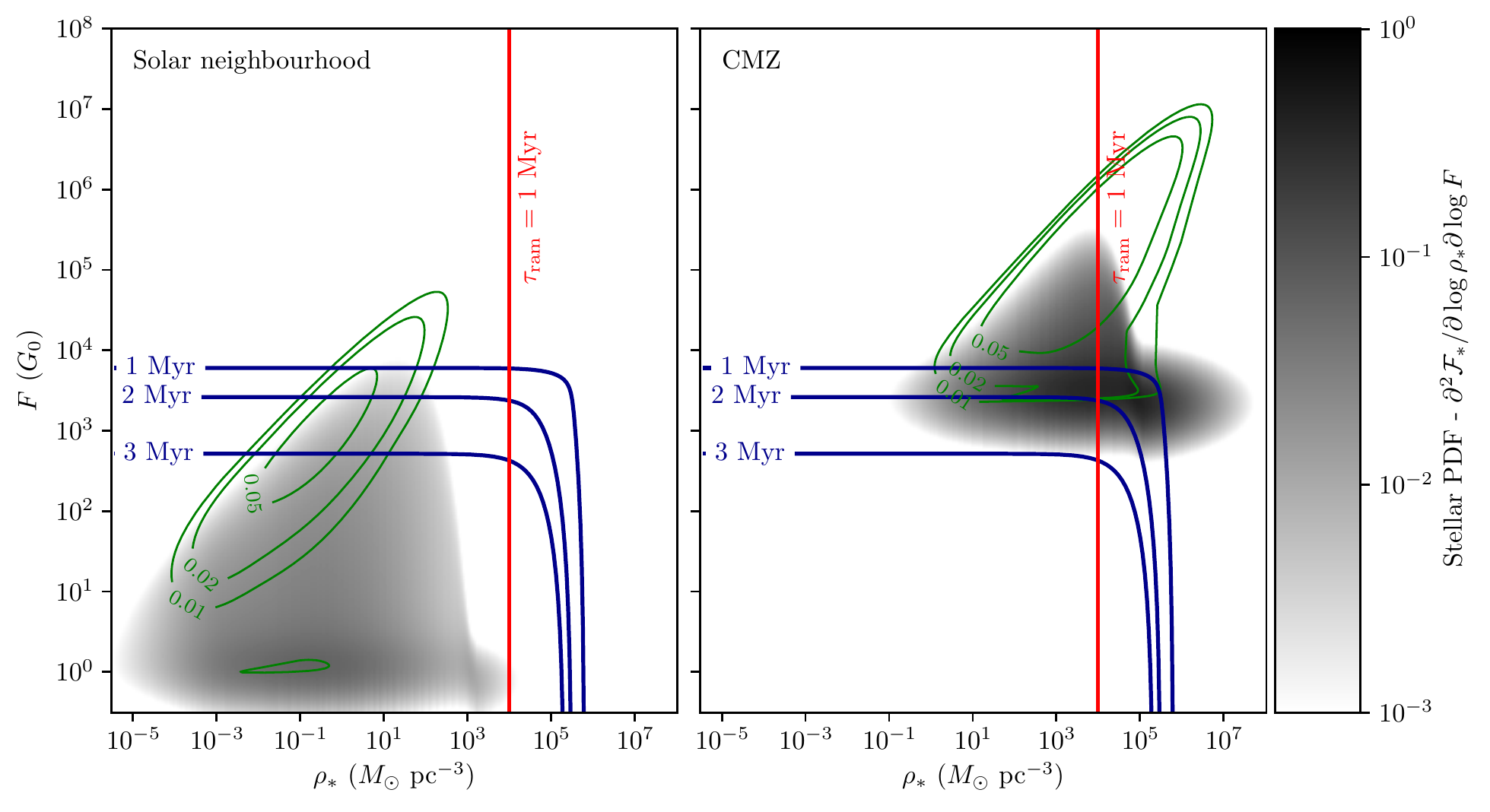}
     \caption{As in Figure~\ref{fig:PDF_Frho}, but including a prescription for the maximal FUV extinction by the ambient gas. The vertical red line marks the stellar density threshold above which ram pressure due to the ambient gas will alter disc evolution on time-scales $\lesssim 1$~Myr for a star with $m_* = 0.5\, M_\odot$ (see the text for details). The green contours are the PDF values without extinction (Figure~\ref{fig:PDF_Frho}) for comparison. We do not indicate the empirical contours in this case, because \citet{Win18b} did not account for extinction in their calculation of the FUV flux. }
     \label{fig:PDF_Frho_ext}
   \end{figure*}

   In Appendix~\ref{sec:FUVext}, we estimate the effect of extinction on the FUV flux distribution by assuming Plummer sphere geometry of the gas within a star-forming region. We then integrate over a radial coordinate defined to be consistent with the total stellar mass to calculate the resulting surface density if the most massive star is at the centre of the region. The gas surface density is then used to calculate the reduction in FUV flux a star experiences. 

   The result of incorporating interstellar extinction into our calculations is shown in Figure~\ref{fig:PDF_Frho_ext}. Our results indicate very high degrees of extinction at high local gas densities $\rho_\mathrm{g}$, and hence we find that many regions where PPDs that would otherwise be dispersed quickly by FUV photons are efficiently shielded during the embedded phase.  Apart from the contours for $\tau_\mathrm{disp}$ due to dynamical encounters and external photoevaporation, we have further indicated a canonical limit above which gas density rapidly alters disc evolution through ram pressure. In the case of the CMZ, the majority of stars that experience large FUV extinction fall into this region, and hence we would expect the ISM to play an important role in PPD evolution prior to gas expulsion. 
   
  As we discuss in Appendix~\ref{sec:FUVext}, the prescription we have implemented for FUV extinction is expected to underestimate the apparent FUV flux experienced by a given star since we have assumed that the local gas density distribution follows a Plummer density profile. This is not the case for a realistic, clumpy gas distributions, which reduce the efficiency of extinction. Hence, the results of our calculations summarised in this section do not offer conclusive answers to the nature of disc evolution during the embedded phase, but rather highlight the importance of the following issues for disc evolution:
  \begin{enumerate}
  \item The time-scale of the embedded phase \citep[e.g.][]{Kru19}. 
  \item The efficiency of extinction during the embedded phase \citep[e.g.][]{Ali19}.
  \item The (statistical) influence of ram pressure stripping on a PPD population as a function of local gas density \citep[e.g.][]{Bat18, Kuf18}. 
  \end{enumerate} 
  In order to fully understand how PPD properties evolve, these three questions must be addressed. Despite these uncertainties, in the case of the CMZ even our calculation for the flux between star-forming regions is sufficient to significantly reduce PPD lifetimes $\tau_\mathrm{disp} \lesssim 3$~Myr. However, we have assumed this lower limit in flux is unaffected by local FUV extinction. We justify this assertion by arguing that, since the field flux $F^\mathrm{f}$ is the sum of contributions from all directions, the clumpiness of the gas distribution makes it likely that it is only reduced by a factor of order unity when averaged over time (dependent on the solid angle subtended by the gas). This assertion requires validation in terms of a realistic treatment of extinction (point ii). For the remainder of this work, we will focus on the lifetimes of discs post-gas expulsion.

\section{Discussion}
\label{sec:discuss}
\subsection{Dispersal time-scale distribution}
\label{sec:disp_dist}
\begin{figure}
       \includegraphics[width=0.46\textwidth]{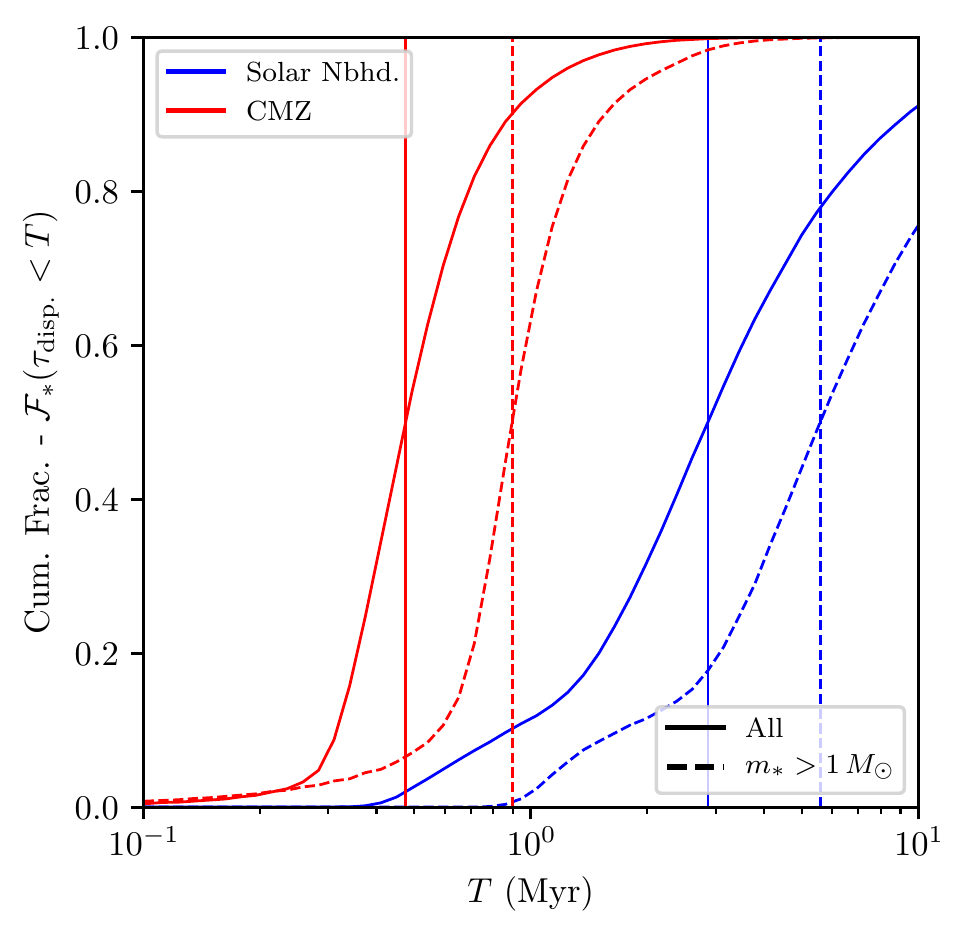}
     \caption{Cumulative fraction of discs with $\tau_\mathrm{disp}<T$ in the solar neighbourhood (blue lines) and CMZ (red lines), as set by external disc dispersal mechanisms. We show the distributions for all discs (solid lines) and discs for which the host star has a mass $>1\, M_\odot$ (dashed lines). The vertical lines of corresponding colour and style mark the median disc lifetimes for each PPD sample. We have again assumed a viscosity parameter  $\alpha = 5.4 \times 10^{-3}$.}
     \label{fig:cdf_tdisp}
\end{figure}
When answering the question of planet formation efficiency within a given environment, it is of major importance to understand the expected distribution of PPD lifetimes. In environments where a large fraction of stars have discs that are quickly dispersed by stellar feedback, we might expect a low planet formation efficiency. For this purpose, we can write the fraction of PPDs with a given lifetime $\tau_\mathrm{disp}$ for fixed $\tau_\mathrm{visc}$:
\begin{equation}
\label{eq:disp_all}
\frac{\partial \mathcal{F}_*}{\partial \tau_\mathrm{disp}} = \int \mathrm{d} F \int \mathrm{d} m_* \frac{\partial \mathcal{F}_*}{\partial F}\frac{\partial \mathcal{F}_*}{\partial m_*} \frac{\partial \mathcal{F}_*}{\partial \rho_*} \left| \frac{\partial \tau_\mathrm{disp}}{\partial \rho_*} \right|^{-1}
\end{equation} where $\partial \mathcal{F}_*/\partial m_* = \xi_*$ is the stellar IMF (we can integrate over flux or stellar density interchangeably here).

The results of this calculation are presented as a cumulative distribution of $\tau_\mathrm{disp}$ for the stellar population in Figure~\ref{fig:cdf_tdisp}. We find that if we consider PPDs around all stars down to $0.08 \, M_\odot$ with our chosen IMF, then we obtain median dispersal time-scales of $2.9$~Myr in the solar neighbourhood and $0.5$~Myr in the CMZ. In both cases these medians are below the characteristic PPD lifetimes for non-photoevaporated populations ($\sim 3$--$10$~Myr). However, if we instead consider only PPDs with host stars above $1 \, M_\odot$, then the median dispersal time-scales increase to $5.6$~Myr in the solar neighbourhood, and $0.9$~Myr in the CMZ. This highlights the large difference between the expected lifetimes of discs around low- and high-mass stars under the influence of external photoevaporation. For all stellar masses, disc lifetimes are suppressed by a factor $\gtrsim 5$ in the CMZ with respect to the solar neighbourhood. This finding has significant consequences for PPD evolution in the central $\sim 250$~pc of the Milky Way, where the time and material available for planet formation is severely reduced by dispersal mechanisms (primarily external photoevaporation). Indeed, for the whole stellar population, $\sim 90\%$ of PPDs are dispersed within $1$~Myr of the destruction of the parent GMC due to external dispersal mechanisms alone.

\subsection{Gas properties \& PPD dispersal}
\begin{figure*}
       \includegraphics[width=0.9\textwidth]{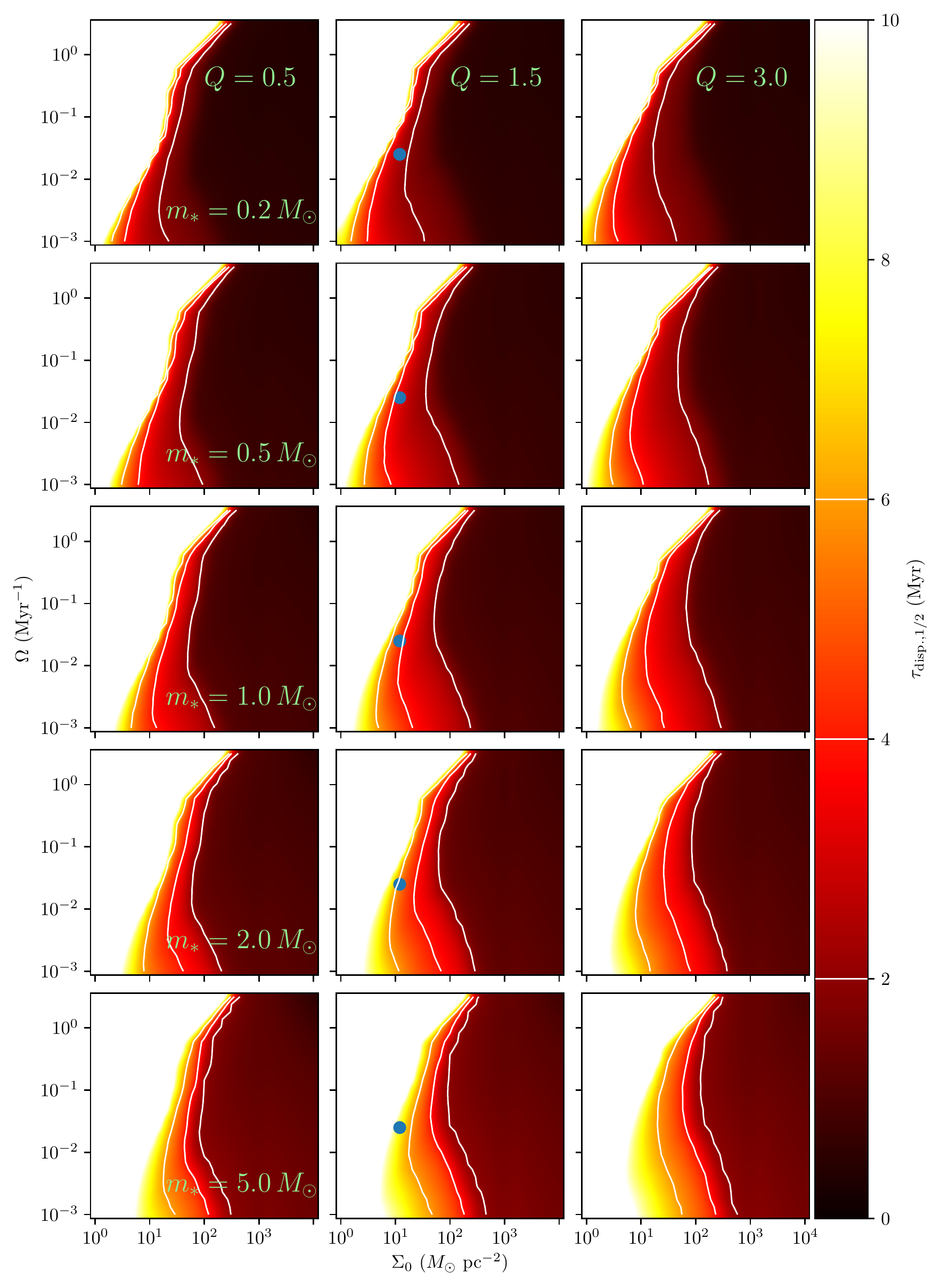}
      \caption{The median dispersal time-scales $\tau_\mathrm{disp,1/2}$ induced by external photoevaporation and dynamical encounters for PPDs around a star of mass $m_*/\msun=\{0.2,0.5,1.0,2.0,5.0\}$ (from top to bottom) as a function of gas surface density $\Sigma_0$ and angular velocity $\Omega$, for Toomre $Q=\{0.5,1,3\}$ (from left to right). The blue circle marks the position of the solar neighbourhood and corresponds to the Sun in the middle panel ($m_*=1\,M_\odot$). Regions of parameter space for which $\tau_\mathrm{disp,1/2}< 10$~Myr exhibit disc lifetimes that are significantly reduced with respect to a PPD evolving in isolation. White contours are placed at $2$, $4$ and $6$~Myr. }
     \label{fig:big_fig}`
\end{figure*}

To explore the parameter space for ISM properties and host stellar mass, we rewrite equation~\ref{eq:disp_all} in terms of a fixed stellar mass:
\begin{equation}
\label{eq:disp_mspec}
\frac{1}{2} =\int_0^{\tau_\mathrm{disp,1/2}}  \! \! \! \!  \! \mathrm{d} \tau_\mathrm{disp} \int \mathrm{d} F \frac{\partial \mathcal{F}_*}{\partial F} \frac{\partial \mathcal{F}_*}{\partial \rho_*} \left| \frac{\partial \tau_\mathrm{disp}}{\partial \rho_*} \right|^{-1} 
\end{equation} to solve numerically for the median dispersal time-scale $\tau_\mathrm{disp,1/2}$. In Figure~\ref{fig:big_fig} we show $\tau_\mathrm{disp,1/2}$ as a function of gas surface density $\Sigma_0$, and angular velocity $\Omega$ within a galactic disc for varying Toomre $Q$, and stellar host mass $m_*$. Most obviously, the time-scale for PPD destruction generally ecreases with increasing $\Sigma_0$. This relationship is simply due to increasing stellar density and maximum mass for a star-forming region with increasing $\Sigma_0$, leading to greater FUV flux. The opposite is true for $Q$, and therefore $\tau_\mathrm{disp}$ increases with increasing $Q$. The increase in $\tau_\mathrm{disp}$ with increasing host mass is due to the greater efficiency of external photoevaporation acting on discs around lower mass stellar hosts, since they have a reduced gravitational potential and therefore smaller gravitational radius within the disc \citep[see][]{Haw18, Haw18b, Win19}.

The dependence of $\tau_\mathrm{disp}$ on $\Omega$ is more complicated, and competing factors dictate the relationship. Firstly, larger $\Omega$ means larger $\rho_0$ (equation~\ref{eq:rho0}), and hence higher densities. However, this also means larger field flux is reduced by greater gas surface density (and therefore extinction). A high angular velocity also restricts the maximum cluster or association mass (equation~\ref{eq:MT}) and therefore reduces the local maximum FUV luminosity, unless $\Omega$ is sufficiently small such that $\langle \tau_\mathrm{fb} \rangle <  \tau_\mathrm{ff, 2D}$ (equations~\ref{eq:tau_ff_2D} and~\ref{eq:fcoll}). In general, high angular velocities decrease the efficiency of externally induced disc dispersal. 

Finally, we find that the position of the solar neighbourhood in the parameter space (marked by a blue dot in the middle panel of Figure~\ref{fig:big_fig}) is approximately at the maximum surface density where the majority of the disc population around stars with $m_*\sim 1\, M_\odot$ do not get significantly depleted by external influences ($\tau_\mathrm{disp}\approx 4$~Myr). This is intriguing because it suggests that the position of the solar system within the galaxy is such that a maximal \textit{number} (not fraction) of stars have PPDs which disperse largely by internal processes (including planet formation). Since the time and material available for planet formation must influence the planets that are capable of forming, we tentatively suggest that the solar neighbourhood is therefore a special region in terms of galactic environments and exoplanet properties (and possibly frequency). Future studies may contextualize this hypothesis in terms of theoretical and observed galactic-scale ISM properties to establish the degree to which the solar neighbourhood is a special case for planet formation. 



\section{Conclusions}
\label{sec:concs}

We have presented the first comprehensive theoretical prescription for linking star formation parameters to PPD dispersal time-scales due to FUV-induced photoevaporation, dynamical encounters and ram pressure stripping. This has numerous applications for assessing the planet formation potential of star-forming regions, and establishing the typical influences on PPD evolution for future investigation. We summarise our main findings as follows:
\begin{enumerate}
\item The solar neighbourhood lies close to the largest ISM surface density for which the majority of the PPD population are not influenced by external dispersal mechanisms. At larger surface densities, PPDs have lifetimes that are significantly shortened by (predominantly) FUV flux. 
\item Due to the higher gas densities in the CMZ, much of the stellar population initially experiences high FUV flux. This results in dispersal time-scales that are a factor $\gtrsim 5$ shorter than those in the solar neighbourhood. Across the entire stellar mass range, we predict that $\sim 90\%$ of PPDs are destroyed within $1$~Myr in the CMZ. Therefore, we expect that planet formation in this region is severely limited in terms of available time and mass. 
\item As found by \citet{Win18b}, external photoevaporation is the dominant mechanism for disc dispersal in the solar neighbourhood, and we find that no stars exist in regions where dynamical encounters can truncate PPDs. Extending this to the CMZ, we find that the time-scale for FUV-induced disc destruction remains shorter than the time-scale for tidal disruption.
\item We estimate an upper limit on the influence of extinction on the FUV flux. Our calculations suggest that PPDs in high density regions ($\rho_* \gtrsim 10^3\,M_\odot$~pc$^{-3}$ in the solar neighbourhood, $\rho_* \gtrsim 10^4\,M_\odot$~pc$^{-3}$ in the CMZ) can be efficiently shielded by ambient gas. In this case dynamical encounters remain insignificant as a depletion mechanism since the ram pressure imposed on a disc population operates on a much shorter time-scale $\tau_\mathrm{ram} \ll \tau_\mathrm{tidal}$ \citep[in agreement with][]{Wij17b}. We therefore conclusively rule out dynamical encounters as the dominant dispersal mechanism in any environment. However, incidental PPD destruction by dynamics remains possible due to the intrinsic stochasticity of this mechanism. For CMZ-like regions, the ram pressure influences PPDs on a short time-scale in all regions where FUV flux is severely reduced by extinction.
\end{enumerate} 

In addition to providing insights into the link between star formation physics and planet formation, our findings also highlight particular questions for future work to answer. For each of the above findings we summarise some such issues:
\begin{enumerate}
\item Is the solar neighbourhood special? Future studies may combine calculations for the number of stars born in a given environment with the expected disc dispersal time-scales we have calculated here. In this way, statistical conclusions can be drawn regarding the significance of the position of the solar neighbourhood in $\Sigma_0$--$\Omega$ space.
\item What is the observed fraction of stars that have discs in the CMZ as a function of age? Early investigations on this topic suggest low disc fractions of a few percent in the Arches cluster \citep{Sto10, Sto15}.
\item How long is the typical viscous time-scale for PPDs? We have assumed a viscous time-scale of $\tau_\mathrm{visc} =1$~Myr for a star of mass $1\, M_\odot$ \citep[broadly consistent with measured accretion rates -- e.g.][]{Man16}. We find that the dispersal time-scale, when dominated by photoevaporation, scales as $\tau_\mathrm{disp} \approx \tau_\mathrm{FUV} \propto  \tau_\mathrm{visc}^{0.7}$, and hence our findings are moderately dependent on the true value of $\tau_\mathrm{visc}$. 
\item What is the influence of ambient gas on disc evolution? This broad topic includes a number of questions regarding both star formation physics and the response of the disc to the ISM. Some of these include: How long is the embedded phase as a function of environment? How efficient is extinction in regions of high gas density? What is the statistical influence of the motion of the dense ISM with respect to a population of PPDs? The first two of these questions can now be addressed systematically with high-resolution imaging of GMC population across the nearby galaxy population \citep{Kru19,Che19}.
\end{enumerate}

Overall, we conclude that building a picture of planet formation predominantly based on PPDs in the solar neighbourhood, or ignoring the dependence of their properties on host stellar mass or the galactic environment will result in a biased understanding of the time and mass available for planet formation over the galactic and cosmological scales relevant for studies of the exoplanet population. The prescription we have presented is a tool for future studies wishing to estimate the variation of PPD properties in diverse environments. Our findings highlight the key issues that need to be addressed in order to further establish the importance of star formation conditions for planet formation.

\section*{Acknowledgements}

We thank the anonymous referee for a considerate report that improved the clarity of this manuscript. We thank Sebastian Trujillo-Gomez for kindly sharing his results, quantifying the  minimum mass of a star forming region, prior to publication. AJW thanks Richard Booth and Cathie Clarke for useful comments and discussion. AJW gratefully acknowledges funding from the European Research Council (ERC) under the European Union's Horizon 2020 research and innovation programme (grant agreement No 681601). AJW gratefully acknowledges support from Sonderforschungsbereich SFB 881 ``The Milky Way System'' (subproject B2) of the German Research Foundation (DFG). JMDK and MC gratefully acknowledge funding from the DFG via an Emmy Noether Research Group (grant number KR4801/1-1). JMDK and BWK gratefully acknowledge from the European Research Council (ERC) under the European Union's Horizon 2020 research and innovation programme via the ERC Starting Grant MUSTANG (grant agreement number 714907). BWK acknowledges funding in the form of a Postdoctoral Research Fellowship from the Alexander von Humboldt Stiftung.




\bibliographystyle{mnras}
\bibliography{truncation} 


\appendix

\section{Short-lived FUV irradiation}
\label{sec:tOB}

\begin{figure}
	\centering
       \includegraphics[width=\columnwidth]{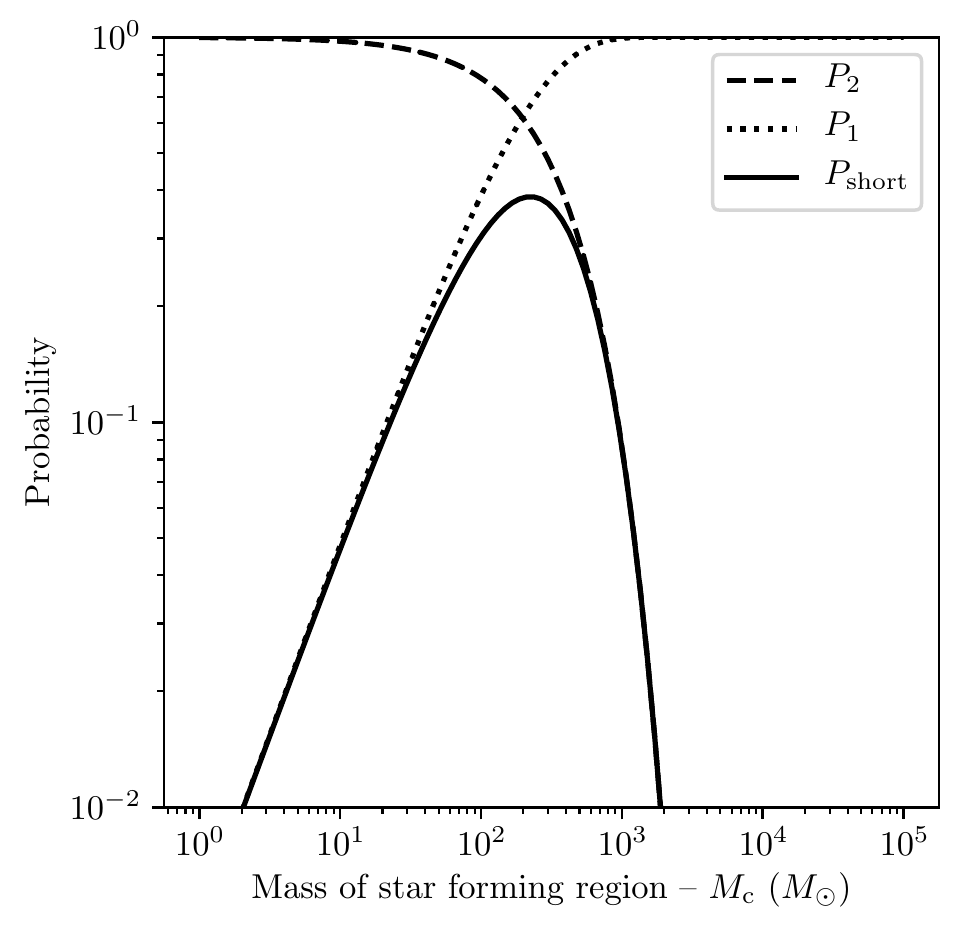}
     \caption{Probability, $P_1$, of a star with a short lifetime ($\lesssim 10$~Myr -- dotted line) and the probability $P_2$, of no star in the range $12$--$16\, M_\odot$ occupying a star forming region of mass $M_\mathrm{c}$ (dashed line). The product of these two probabilities $P_\mathrm{short}$ is the probability of a star forming environment being strongly irradiated for a period $<10$~Myr (equation~\ref{eq:pshort} -- solid line).}
     \label{fig:pshort}
\end{figure}
We have assumed in our models that massive stars in a given star forming region do not reach the end of their lifetime before PPDs are dispersed. In this appendix we explore this assumption. The main sequence lifetime of stars can be approximated:
$$
\frac{\tau_\mathrm{MS}}{10^4 \,  \mathrm{Myr}} \sim \left( \frac{m_*}{M_\odot}\right)^{-2.5}.
$$ For the strongest FUV environments where $\tau_\mathrm{FUV} \lesssim 1$~Myr, our approximation is reasonable since even the most massive stars survive over this time-scale. For stars of stellar mass $m_* \gtrsim 12\, M_\odot \equiv M_\mathrm{FUV}$, the FUV luminosity is within approximately an order of magnitude of the most massive stars \citep[e.g.][]{Win18b}. For stellar masses $m_* \gtrsim 16 \, M_\odot \equiv M_\mathrm{short}$, $\tau_\mathrm{MS}\lesssim 10$~Myr. 

To approximate the frequency of systems where PPDs can be strongly irradiated for short periods, we estimate the probability $P_\mathrm{short}$ that a short-lived star with $m_*>M_\mathrm{short}$ exists in a star forming region (with probability $P_1$), but there is no star with a mass in the range $ M_\mathrm{FUV}< m_*< M_\mathrm{short}$ (with probability $P_2$). If the latter condition is not met, then even when the most massive star in the region reaches the end of its lifetime, then the drop in the FUV luminosity of the most massive remaining star will be less than an order of magnitude. The fraction of stars between masses $M_1$ and $M_2$ is:
\begin{equation}
\Xi_{M_1}^{M_2} = \int_{M_1}^{M_2} \xi_* \mathrm{d} m_*
\end{equation} where $\xi_* $ is the normalised IMF (equation~\ref{eq:imf}). Then we have:
\begin{equation}
\label{eq:pshort}
    P_\mathrm{short} = \underbrace{\left[ 1 - \left( 1 - \Xi_{M_\mathrm{short}}^\infty\right)^{N_\mathrm{c}}\right] }_{P_1} \cdot \underbrace{\left( 1 - \Xi^{M_\mathrm{short}}_{M_\mathrm{FUV}}\right)^{N_\mathrm{c}}}_{P_2} 
\end{equation} where the number of stars $N_*$ in the star forming region is $N_\mathrm{c}=M_\mathrm{c}/\langle m_* \rangle \approx M_\mathrm{c}/0.5 \, M_\odot$.

The result of the calculation of $P_1$ and $P_2$ as a function of the stellar mass of the star forming region is shown in Figure~\ref{fig:pshort}. In the range $\sim 30$--$1000\, M_\odot$, where $P_\mathrm{short}$ is maximised, we find that there is a $10$--$40$ percent chance of a short lived period of strong exposure to FUV flux. This possibility warrants further exploration in future investigations, however we do not do so here. We justify our simplification in that the regions of the highest FUV flux exhibit disc lifetimes $\ll 10$~Myr, so our estimates in this section are an upper limit on the significance of short main sequence lifetimes.

\section{FUV extinction}
\label{sec:FUVext}
\subsection{Modified flux distribution}
At early times, the presence of ambient gas causes intra-cluster extinction in the FUV; we wish to evaluate its influence on the flux PDF at fixed $x$. This is dependent on the effective local gas surface density $\Sigma_\mathrm{eff}$ between a given star and FUV source. We define the corresponding surface overdensity $\chi \equiv \Sigma_\mathrm{eff}/\Sigma_0$.  In order to proceed, we assume that the local extinction does not influence the flux in the field, which remains the floor of the distribution of $F$. Then the ratio of the extincted flux to the local mean flux is $\psi^\mathrm{ext}_0 \equiv F^\mathrm{ext}_0/F_0^\mathrm{HM}$ is
\begin{equation}
\label{eq:psi_ext}
\psi^\mathrm{ext}_0  =   e^{- C_\mathrm{ext}\chi}\Lambda+ \psi^\mathrm{f}_0.
\end{equation} As before, we can immediately evaluate the PDF for $\psi_0^\mathrm{ext}$ at certain limits. Equation~\ref{eq:dpdpsi0_psi0f} applies here as before, as does equation~\ref{eq:dpdpsi0_upper}. However, $\psi_0^\mathrm{ext}$ is now determined by $\chi$ as well as $x$ and $\Lambda$. We must therefore evaluate the PDF for $\chi$.


\subsection{Effective surface density}
\subsubsection{Gas density profile}
To evaluate the appropriate surface density, we are required to make assumptions about the geometry of the system. This involves introducing an additional parameter, describing the relative position in a local environment such that we can link $\rho_\mathrm{g}$ to $\Sigma_\mathrm{eff}$. We define a radial coordinate $r$ within a star-forming region of scale radius $a$, and the relative radius $\gamma \equiv r/a$. \citet{Fat08} define a \citet{Her90} density profile to calculate the apparent surface density. However, this form implies an infinite central density, which is inconsistent with our assumption that gas density is lognormally distributed. We instead choose a Plummer density profile: 
\begin{equation}
\label{eq:gamma_rho}
\rho_\mathrm{g} =  \frac{\rho_{\mathrm{c}}}{(1+\gamma^2)^{5/2}}.
\end{equation}The local overdensity in the centre $x_\mathrm{c} \equiv \rho_{\mathrm{c}}/ \rho_0$ is a bijective function of $\gamma>0$ for  $x<x_\mathrm{c}$, and we assume the same lognormal PDF as for $x$ truncated below this value. The corresponding PDF for $\gamma$ at fixed $x$ is 
\begin{equation}
\label{eq:dpdgamma}
\frac{\partial \mathcal{F}_*}{\partial \gamma} = \frac{\partial \mathcal{F}_*}{\partial x_\mathrm{c}}  \frac{\partial x_\mathrm{c}}{\partial \gamma} \propto \gamma \left(1+\gamma^2\right)^{4}\frac{\partial p}{\partial x_\mathrm{c}},
\end{equation} where $\partial p/\partial x_\mathrm{c}  \propto \partial p/\partial x$ for $x< x_\mathrm{c}$ and vanishes otherwise. As a sanity check, we consider the functional dependence on the distribution of $\gamma$ for small and large $x$. Since the overdensity PDF is lognormal, for large $x$ then as $x$ increases the (negative) slope of the PDF for $x_\mathrm{c}$ also increases. Therefore, at large $x$ we preferentially find small $\gamma$. This is exactly what we would expect since extremely high densities should be rare at large radii. Similarly, for small $x$ a large value of $\gamma$ is favoured. For a given $\gamma$, we can also calculate the corresponding $a$ such that the total gas mass is $\phi M_\mathrm{crit}/\epsilon$:
\begin{equation}
\label{eq:Rc}
a = \left( \frac{3 M_\mathrm{crit}}{4\pi \epsilon \rho_0 x_\mathrm{c} } \right)^{1/3} = \left( \frac{3 M_\mathrm{crit}}{4\pi \epsilon\rho_0 } \right)^{1/3} x^{-1/3} \phi^{1/3}{(1+\gamma^2)^{-5/6}}.
\end{equation} Thus we define a density profile that is self-consistent with a given $x$, $\phi$.

\subsubsection{Ionisation}

\begin{figure}
	\centering
       \includegraphics[width=\columnwidth]{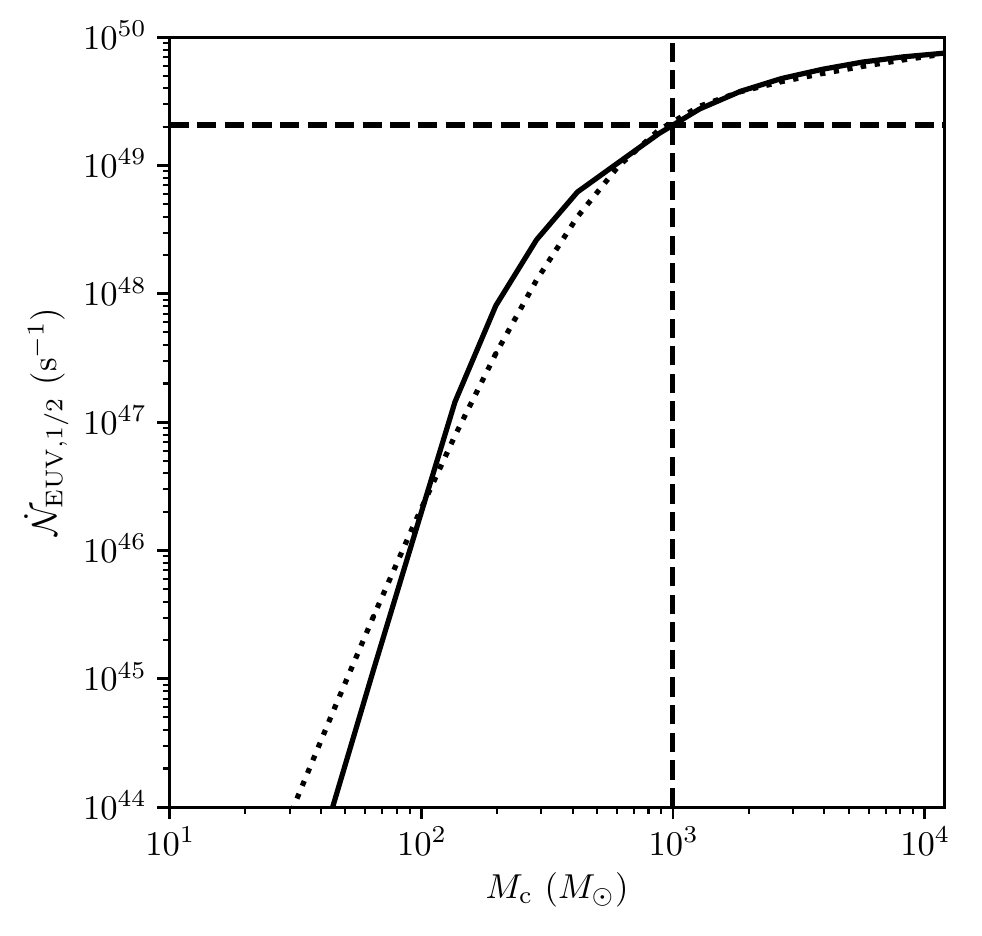}
     \caption{As in Figure~\ref{fig:Lhalf} but for the number of EUV counts $\dot{\mathcal{N}}_{\mathrm{EUV},1/2}$. The solid line is calculated directly from random drawing and the stellar atmosphere models used in this work, while the dotted line follows our analytic approximation, equation \ref{eq:Theta}.  The vertical dashed line is at $M_\mathrm{crit}$ and the corresponding number of counts $\dot{\mathcal{N}}_{\mathrm{EUV,crit}} =  2.07 \times 10^{49}$~s$^{-1}$ is shown as a horizontal dashed line.}
     \label{fig:EUVhalf}
\end{figure}

Having defined our local density profile, we integrate over the relevant range to establish the effective surface density. When a massive star occupies the central region of a given environment then we would expect material within a certain radius to be ionised (and therefore optically thin to FUV photons). This size scale is initially given by the \citet{Str39} radius:
\begin{equation}
\label{eq:RS}
R_\mathrm{S} \approx \left( \frac{3 \dot{\mathcal{N}}_\mathrm{LyC} m_\mathrm{p}^2}{4\pi \alpha_\mathrm{B} \rho_\mathrm{c}^2} \right)^{1/3}  = \left( \frac{3 \dot{\mathcal{N}}_\mathrm{LyC} m_\mathrm{p}^2}{4\pi \alpha_\mathrm{B} \rho_0^2} \right)^{1/3} x^{-2/3} (1+\gamma^2)^{-5/3} 
\end{equation} 
where $\dot{\mathcal{N}}_\mathrm{LyC}$ is the number of ionising (Lyman continuum) photons emmitted by the central source per unit time, $\alpha_\mathrm{B} \approx 2.7 \times 10^{-13}$~cm$^3$~s$^{-1}$ is the recombination coefficient assuming a temperature $\sim 10^4$~K for the ionised gas. For convenience we have approximated a constant local density for $r<R_\mathrm{S}$, which holds for if $R_\mathrm{S} \lesssim a$. This is true for large $x$, where we will find that extinction is significant.

We assume EUV photons dominate ionisation and define the median number of EUV counts from the most massive star $\dot{\mathcal{N}}_{\mathrm{EUV},1/2}(\phi)$. For this we define a fitting formula:
\begin{equation}
\label{eq:Theta}
\Theta(\phi) \equiv \frac{\dot{\mathcal{N}}_{\mathrm{EUV},1/2}}{\dot{\mathcal{N}}_{\mathrm{EUV, crit}}} \approx \left\{ 1-e^{ - \delta_1  \phi  }\right\}^{\delta_2} \ln \left( 1+ \delta_1 \phi \right)
\end{equation} where $\dot{\mathcal{N}}_{\mathrm{EUV, crit}} = 2.07 \times 10^{49}$~s$^{-1}$, and we find $\delta_1 = 2.9$, $\delta_2 = 4.0$. This expression is compared to the direct calculation from the adopted stellar atmosphere models in Figure~\ref{fig:EUVhalf}. 

Combining equations \ref{eq:RS} and \ref{eq:Theta}, we have:
\begin{multline}
\label{eq:gammaS}
\gamma_\mathrm{S} \equiv \frac{R_\mathrm{S}}{a}\\ = \left( \frac{ \dot{\mathcal{N}}_\mathrm{EUV, crit}  m_\mathrm{p}^2}{\alpha_\mathrm{B} M_\mathrm{crit} \rho_0}\right)^{1/3} 
 \epsilon ^{1/3} x^{-1/3}\phi^{-1/3} \left(1+\gamma^2\right)^{-5/6}\Theta^{1/3}, 
\end{multline}in dimensionless quantities. Evaluating the prefactor yields:
\begin{equation}
\left( \frac{ \dot{\mathcal{N}}_\mathrm{EUV, crit}  m_\mathrm{p}^2}{\alpha_\mathrm{B} M_\mathrm{crit} \rho_0}\right)^{1/3}  \approx \left(\frac{\rho_0}{1.6 \, M_\odot \, \mathrm{pc}^{-3}}  \right)^{-1/3}.
\end{equation}

\subsubsection{Effective surface density PDF outside Str\"{o}mgren radius}

We assume that $F$ is dominated by sources at the center of the density profile. Assuming spherical geometry, then the effective surface density is that of a spherical shell and we have:
\begin{equation}
\label{eq:sigma_eff}
\Sigma_\mathrm{eff} =  a\int_{\gamma_\mathrm{S}}^{\gamma} \! \frac{(1-\epsilon)\rho_\mathrm{c}}{(1+\tilde{\gamma}^2)^{5/2}} \, \mathrm{d}\tilde{\gamma}
\end{equation}By making the simplifying assumption that the SFE is approximately constant over the region such that $\epsilon = \epsilon(x) \neq \epsilon(\gamma)$, we can evaluate equation~\ref{eq:sigma_eff} in terms of dimensionless parameters:
\begin{multline}
\label{eq:chi}
\chi_1 \approx \frac{(1-\epsilon)}{3 \epsilon^{1/3}} \left( \frac{3 M_\mathrm{crit} \rho_0^2}{4\pi\Sigma_0^3} \right)^{1/3} x^{2/3} \phi^{1/3}(1+\gamma)^{5/3}  \times \\
\left\{\frac{\gamma(2\gamma^2+3)}{(1+\gamma^2)^{3/2} }- \frac{\gamma_\mathrm{S}(2\gamma_\mathrm{S}^2+3)}{(1+\gamma_\mathrm{S}^2)^{3/2} } \right\} ,
\end{multline} where we have defined $\chi_1 \equiv \chi(\gamma>\gamma_\mathrm{S})$, since $\chi (\gamma<\gamma_\mathrm{S}) = 0$.

\begin{figure}
  \includegraphics[width=0.46\textwidth]{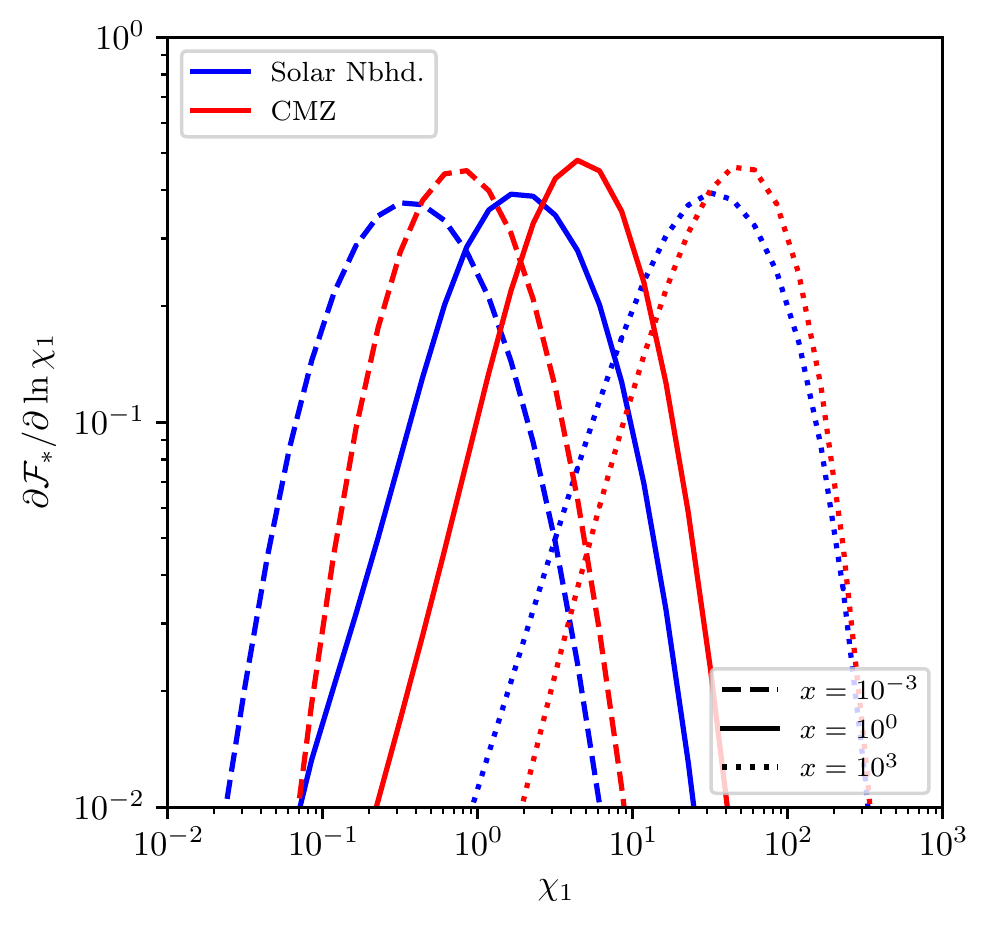}
     \caption{PDF of $\chi_1$ (i.e.\ $\chi$ such that we assume $\chi>0$) for varying overdensity $x$ in the solar neighbourhood (blue lines) and CMZ (red lines). The value of $\chi_1$ is the normalised effective surface density experienced by a given star in the direction of the centre of the star-forming region during the embedded phase. }
     \label{fig:dpdchi1}
\end{figure}

\begin{figure}
       \includegraphics[width=0.46\textwidth]{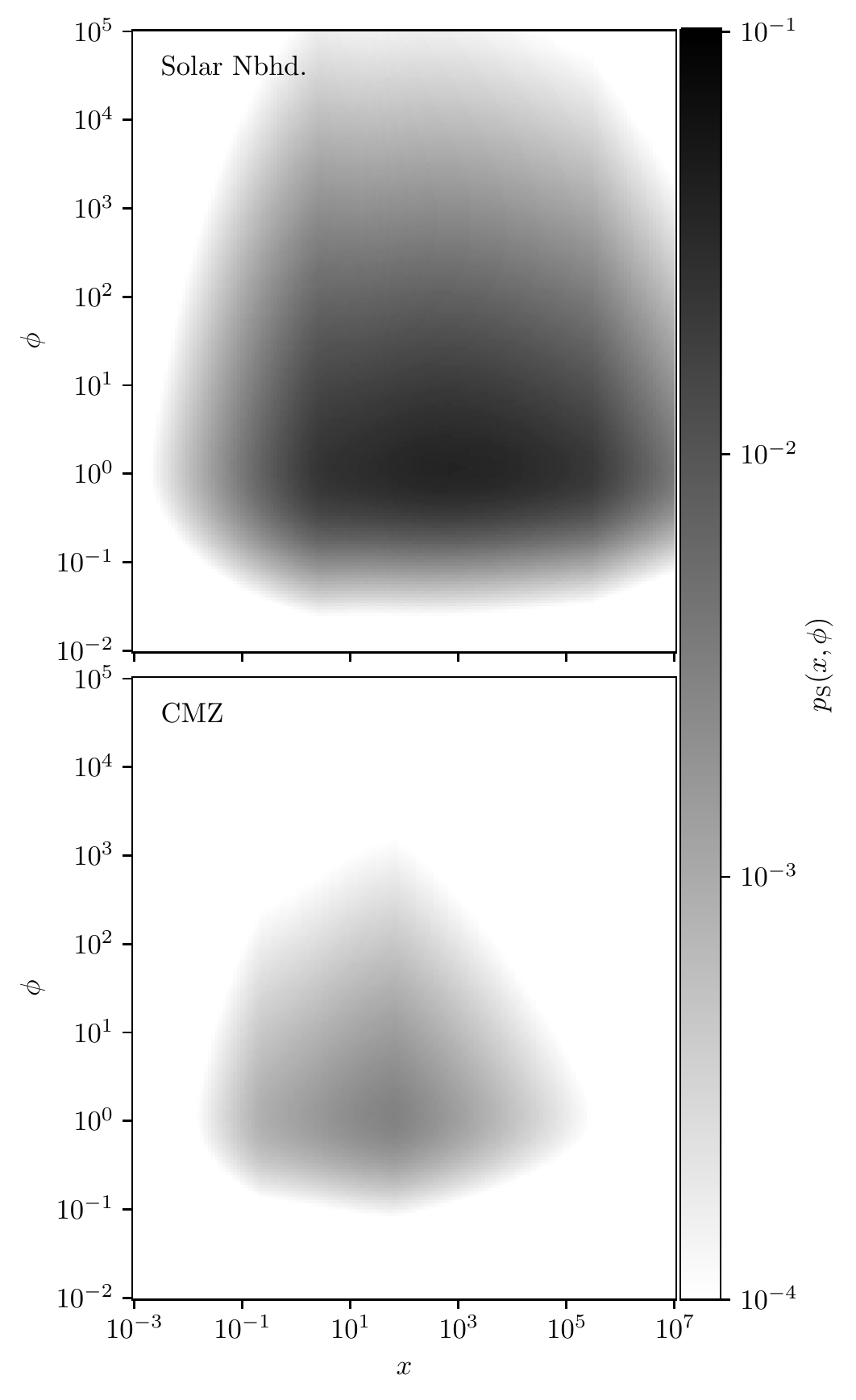}
     \caption{Probability $p_\mathrm{S}$ of finding a given star within the Str\"{o}mgren radius ($\gamma<\gamma_\mathrm{S}$) as a function of the stellar mass $\phi$ of the star-forming region and overdensity $x$ for the solar neighbourhood (top panel) and CMZ (bottom panel). This is equivalent to the probability that a star has $\chi=0$, and does not experience significant extinction of FUV photons from neighbouring stars. }
     \label{fig:pchi0}
   \end{figure}
  
Now we have a definition for $\chi$, we can use the PDF for $\gamma$ (equation~\ref{eq:dpdgamma}) and $\phi$ (equation~\ref{eq:MCMF}) to calculate the corresponding PDF for $\chi$ at a fixed $x$. However, since there is a non-zero probability that $\chi=0$, we must separately consider the regions inside and outside the Str\"{o}mgren radius. We first evaluate the PDF of $\chi_1$ (that is, assuming $\chi>0$ -- a star outside the Str\"{o}mgren radius): 
\begin{equation}
\label{eq:dpdchi1}
\frac{\partial \mathcal{F}_*}{\partial \chi_1} = \int \mathrm{d} \phi \frac{\partial \mathcal{F}_*}{\partial \phi}  \frac{\partial \mathcal{F}_*}{\partial \gamma} \left|\frac{\partial \chi_1} {\partial \gamma}\right|^{-1},
\end{equation}Here it is necessary to evaluate $\gamma(\phi, \chi_1)$  numerically. The result is shown in Figure~\ref{fig:dpdchi1} for solar neighbourhood- and CMZ-like regions. We find that the effective surface density experienced by a given star increases with local gas density $x$, as expected. We therefore expect regions of high overdensity to be severely influenced by extinction. However, we must also consider the fact that stars at high density are more likely to be found towards the centre of the region, and therefore to occupy the Str\"{o}mgren sphere (hence $\chi=0$). It is necessary to explore the possibility that this influences our results.

  \subsubsection{Fraction of stars born within the Str\"{o}mgren radius}
  
The probability that $\chi=0$ is equivalent to the probability that a star is found inside a radius $\gamma_\mathrm{S}$. This can be written:
\begin{equation}
\label{eq:pchi_small}
p_\mathrm{S}(\phi, x) \equiv \mathcal{F}_*( \chi=0; \phi, x) = \int_{\gamma<\gamma_\mathrm{S}}  \! \! \mathrm{d} \gamma \frac{\partial \mathcal{F}_*}{\partial \gamma}  ,
\end{equation} where the region $\gamma<\gamma_\mathrm{S}$ is defined numerically for a fixed $\phi$, $x$. Equation~\ref{eq:pchi_small} is evaluated in Figure~\ref{fig:pchi0}, from which we find that the probability of finding a star within a Str\"{o}mgren radius is small ($\ll 10 \%$) throughout the parameter space, especially for high $\rho_0$ environments. This is intuitively true from equation~\ref{eq:gammaS}; in the limit of large $\rho_0$, $x$, $\phi$, we have small $\gamma_\mathrm{S}$, and hence a small $p_\mathrm{S}$. Since the contribution to the PDF from stars with $\gamma < \gamma_\mathrm{S}$ is small, we have $\partial \mathcal{F}_*/\partial \chi_1 \approx \partial \mathcal{F}_* / \partial \chi$ and we are free to limit our consideration to the distribution of $\chi_1>0$ in calculation of the PDF for $\psi_0^\mathrm{ext}$. While we are here interested in initial conditions, it should also be noted that physically this radius expands over time (see Section~\ref{sec:PDF_exflux} below).

\subsection{PDF for extincted FUV flux}
\label{sec:PDF_exflux}
As a result of the above analysis, we can now simply write the PDF for $\psi_0^\mathrm{ext}$:
\begin{equation}
\label{eq:psi0_ext_PDF}
\frac{\partial \mathcal{F}_*} {\partial \psi_0^\mathrm{ext}} \approx   \int_{\delta \chi}^\infty \! \! \mathrm{d} \chi_1 \,  \frac{\partial \mathcal{F}_*}{\partial \chi_1}  \frac{\partial \mathcal{F}_*}{\partial \phi} \left|\frac{\partial \psi_0^\mathrm{ext}} {\partial \phi}\right|^{-1}  
\end{equation} for some sufficiently small $\delta \chi$.  Equation~\ref{eq:psi0_ext_PDF} is the PDF for the flux in the embedded phase of the cluster or association (at fixed $x$), and can be compared to the non-extincted PDF (equation~\ref{eq:psi0_PDF}) to estimate the role of gas with regards to stellar birth environment at early times. 

We stress that this formulation gives an upper limit to the extinction experienced within a given environment. We have neglected the fact that realistically we would expect a clumpy density distribution, which can considerably reduce the influence of extinction \citep{Ali19}. Additionally, we have established the Str\"{o}mgren radius by assuming a constant central density, and the EUV luminosity of the single most massive star. In the case of a steep density profile, or multiple ionising sources, this will be an underestimate. Nor have we considered the rate of expansion of such an ionised region over time $t$; physically the radius of the ionised region scales with $(c_\mathrm{s} t/R_\mathrm{S})^{4/7}$ \citep[e.g.][]{Spi78, Bis15}. For these reasons, the true FUV flux experienced by a star is likely to be larger than the estimate we establish here.

\bsp	
\label{lastpage}
\end{document}